\documentclass[journal=jpccck,manuscript=article,layout=onecolumn]{achemso}
\usepackage{chemformula} 
\usepackage[T1]{fontenc} 
\usepackage{hyperref}
\usepackage{comment}
\usepackage{amsmath}
\usepackage{stfloats}
\usepackage{xcolor}
\usepackage{graphicx}

\author{J. P. Dadario Pereira}
\affiliation{Applied Physics Department, ’Gleb Wataghin’ Institute of Physics, State University of Campinas, Campinas,SP, 13083-970, Brazil}

\author{Raphael Tromer}
\affiliation{University of Brasília, Institute of Physics, Brasília, Federal District, Brazil}

\author{Luiz A. Ribeiro Junior}
\affiliation{Computational Materials Laboratory, LCCMat, Institute of Physics, University of Bras\'ilia, 70910-900, Bras\'ilia, Federal District, Brazil}

\author{Douglas S. Galvao}
\affiliation{Applied Physics Department, ’Gleb Wataghin’ Institute of Physics, State University of Campinas, Campinas,SP, 13083-970, Brazil}

\title[An \textsf{achemso} demo]
  {Hopping-Mediated Charge Transport in Graphene Beyond the Ballistic Regime}

\begin{document}


\begin{abstract}
  
We introduce a trajectory-resolved framework for charge transport in graphene and similar two-dimensional carbon systems, functioning beyond the idealized ballistic and fully coherent regimes. Transport is modeled using kinetic Monte Carlo hopping dynamics on a predefined atomic lattice, enabling concurrent analysis of disorder, thermal activation, and external fields. We can obtain the current and the effective transmittance directly from random carrier paths, without recourse to phenomenological transport coefficients. We use this method to measure how bias voltage (0–0.10 V), temperature (300–900 K), magnetic field (0–10 T), in-plane strain (2–10\%, uniaxial and biaxial), and vacancy concentration (0–10\%) change the paths that electrons take through graphene networks. For pristine graphene, the response is almost ohmic over the bias window we examined. At 0.10 V, the current is about 7–8 $\mu$A, the effective transmittance goes up to about 0.98–1.00, and the conductance is about (5.8–7.8) $\times 10^{-5}$ S, depending on the direction. Vacancies progressively suppress transport: at 10\% vacancy, the transmittance can drop to ~0.45–0.75 (depending on bias and direction), and the current is significantly reduced compared to the pristine case. Increasing the temperature (from 300 K to 900 K) speeds up hopping kinetics and partially restores transport, but it can't fully compensate for the loss of connectivity that occurs when many defects are present. External magnetic fields cause additional suppression, with the largest drops occurring in more disordered networks and at high fields (up to 10 T). In general, the framework brings together current, transmittance, and field- and defect-dependent transport regimes into one clear computational scheme for realistic two-dimensional carbon materials. It also allows the determination of diffusion coefficients and effective mobilities from carrier displacements and transit times. 
\end{abstract}

\section{Introduction}

In condensed matter physics and nanoelectronics, it is still challenging to understand how charge moves through low-dimensional carbon-based materials~\cite{sp2}. Graphene has long been a model system because it has no gap in its electronic structure, its carriers move quickly, and it has an almost perfect ohmic response in the crystalline and weakly disordered limit~\cite{castro2009electronic,ElectronicTwoGraph}. Nevertheless, in practical experimental settings, graphene is rarely entirely unblemished. Structural defects, vacancies, lattice distortions, substrate interactions, and external fields inherently induce disorder, disrupting translational symmetry and gradually diminishing extended electronic states~\cite{RevDefcts,yang2018structure,mucciolo,Lherbier2012,ugeda2010missing}.

As disorder rises, transport in graphene-derived systems can significantly diverge from coherent band conduction, transitioning into regimes characterized by localization, percolation, and thermally facilitated hopping between spatially localized states~\cite{qiu2013hopping,PhysRev.120.745,shklovskii1984variable,de2024strain,da2015impurity}. It is still challenging to fit this crossover into a single theoretical framework. Fully coherent quantum transport methods that use non-equilibrium Green's functions (NEGF) work very well for short, clean devices~\cite{camsari2022nonequilibrium,ryndyk2016theory}, but they don't work as well in large systems, when there is a lot of disorder, or when the temperature is high~\cite{polanco2021nonequilibrium}. On the other hand, classical drift-diffusion models don't have the microscopic detail needed to link changes in atomic structure to new transport behavior~\cite{degond2005quantum}. These constraints drive the advancement of intermediate-scale methodologies that can integrate atomistic structure, electronic localization, and non-equilibrium transport.

In this work, we adopt a microscopic hopping-based framework to investigate charge transport in graphene-based systems under the combined influence of structural disorder, mechanical strain, temperature, and external magnetic fields. The approach is explicitly designed to operate in regimes where transport is governed by spatial connectivity and localization of electronic states, rather than by phase-coherent band propagation. Pristine graphene is therefore not treated as an ideal Dirac conductor within this framework, but rather as a limiting case of maximal lattice connectivity against which localization-driven deviations induced by disorder can be systematically quantified. Our model considers finite temperatures starting from room temperature; in this regime, collisions with phonons—representing lattice vibrations—destroy quantum coherence, so transport ceases to be ballistic and becomes diffusive, which can be approximated by particle hopping \cite{hwang2008acoustic}.

While graphene is renowned for its ballistic transport properties over micrometer scales at low temperatures, such a regime is fundamentally altered in the presence of significant thermal fluctuations and structural disorder \cite{Morozov2008, Bolotin2008}. At temperatures above 300 K, electron-phonon scattering becomes a dominant factor, significantly reducing the phase coherence length \cite{hwang2008acoustic}. Furthermore, the introduction of vacancies and structural defects introduces a strong scattering environment where the mean free path becomes comparable to the lattice constant \cite{Chen2009}. In this high-disorder and high-temperature limit, the electronic states become increasingly localized, and the transport mechanism transitions from a coherent wave-like propagation to a stochastic, hopping-mediated process \cite{Lherbier2008}.

A central feature of the methodology is the explicit separation between electronic structure characterization and transport modeling. Electronic structure calculations based on a distance-dependent tight-binding model are employed solely to extract an effective energy scale associated with disorder- or strain-induced suppression of extended states. This scale, referred to here as an effective transport gap or mobility gap \cite{shklovskii1984variable}, is not interpreted as the intrinsic band gap of pristine graphene. Instead, it provides a physically motivated parameter controlling wavefunction localization and hopping dynamics within the transport model. In this way, structural perturbations such as vacancies and lattice strain influence transport only through their impact on localization and connectivity, preserving a clear causal link between atomic-scale modifications and macroscopic current response.

Charge transport is simulated using a kinetic Monte Carlo description of incoherent hopping between localized states \cite{bortz1975new}. This formalism captures the essential physics of thermally activated transport, bias-induced energy tilting, and magnetic-field-induced localization \cite{abrahams1979scaling}, while remaining computationally efficient enough to enable systematic exploration of wide parameter spaces. The model naturally reproduces ohmic current--voltage characteristics in highly connected networks and predicts strong deviations from linear response as localization is progressively enhanced \cite{shklovskii1984variable}. Importantly, the kinetic Monte Carlo approach is employed to capture the percolative nature of charge carriers navigating this disordered landscape, providing a more realistic description of graphene devices operating under ambient or extreme conditions.

The main goal of this study is not to replicate graphene transport in the clean limit but to clarify how deviations from ideality, driven by disorder, strain, and external fields, encourage localization. We systematically alter vacancy concentration, temperature, magnetic field strength, and applied strain to elucidate the mechanisms that facilitate the transition from graphene-like transport to regimes characterized by activated hopping and magnetolocalization \cite{abrahams1979scaling, bergmann1984weak, chen2010magnetoresistance, gorbachev2007weak}. The framework and insights developed here can be applied to other two-dimensional materials and nanostructures beyond graphene, where disorder and lattice deformations are important for determining how electrons move.

\section{Methodology}

Electronic transport is described within a microscopic framework tailored to low-dimensional systems in which charge carriers occupy spatially localized states and conduction proceeds via thermally assisted hopping~\cite{PhysRev.120.745,shklovskii1984variable}. This regime is particularly relevant for atomically thin materials exhibiting structural disorder, functionalization, grain boundaries, or reduced coherence lengths, where fully coherent quantum transport and semiclassical drift--diffusion models become inadequate~\cite{datta1997electronic,peres2010colloquium}.

\subsection{Atomistic structure and connectivity}

Atomic geometries are obtained directly from crystallographic information files (CIF) ~\cite{shi2021high}, ensuring faithful representation of the underlying lattice. When required, the primitive cell is replicated along the in-plane directions to construct extended simulation domains. The atomic coordinates are rigidly rotated such that one lattice vector aligns with the chosen transport direction, allowing an unambiguous definition of source and drain regions.

Each atom is treated as a localized transport site characterized by its planar coordinates $\mathbf{r}_i = (x_i, y_i)$. Connectivity between sites is defined geometrically: two sites $i$ and $j$ are considered neighbors if their separation satisfies
\begin{equation}
r_{ij} = |\mathbf{r}_j - \mathbf{r}_i| \le r_c,
\end{equation}
where $r_c$ is a cutoff distance chosen to reflect physically meaningful orbital overlap~\cite{papaconstantopoulos1986handbook,slater1954simplified}. This procedure preserves the local topology of the atomic lattice without introducing artificial long-range connections.

\subsection{Electrostatic bias and site energies}

An external bias voltage $V$ applied between source and drain is incorporated through a linear electrostatic potential drop along the transport direction. The on-site energy of each localized state is given by
\begin{equation}
E_i(V) = -eV \, \frac{u_i - u_{\mathrm{min}}}{L},
\end{equation}
where $u_i$ is the coordinate of site $i$ along the transport axis, $u_{\mathrm{min}}$ is its minimum value, $L$ is the device length, and $e$ is the elementary charge. This prescription corresponds to a uniform electric field driving carriers from source to drain.

\subsection{Hopping transition rates}

Charge transport between neighboring sites is modeled as an incoherent hopping process governed by Miller--Abrahams transition rates~\cite{PhysRev.120.745}. The hopping rate from site $i$ to site $j$ is defined as
\begin{equation}
\Gamma_{i \rightarrow j} =
\nu_0 \, \frac{1}{m_{\mathrm{eff}}}
\exp\!\left(-\frac{2 r_{ij}}{\xi}\right)
\times
\begin{cases}
\exp\!\left(-\dfrac{E_j - E_i}{k_{\mathrm{B}} T}\right), & E_j > E_i, \\
1, & E_j \le E_i,
\end{cases}
\end{equation}
where $\nu_0$ is an attempt frequency, $k_{\mathrm{B}}$ is Boltzmann's constant, $T$ is the temperature, and $\xi$ is the effective localization length controlling the spatial decay of electronic wavefunctions.

\subsection{Effective mass and localization length}

The suppression of carrier mobility associated with an effective transport gap $E_g$ is incorporated through a phenomenological effective mass \cite{Grundmann2021},
\begin{equation}
m_{\mathrm{eff}} = 1 + 2E_g,
\end{equation}
with $E_g$ expressed in electronvolts. This form ensures a monotonic reduction of hopping probability with increasing localization energy scale and should be interpreted as an effective transport parameter rather than a band mass.

The localization length is renormalized according to
\begin{equation}
\xi_g = \frac{\xi_0}{1 + \alpha E_g},
\end{equation}
where $\xi_0$ is the bare localization length and $\alpha$ controls gap-induced localization. Thermal effects are included as
\begin{equation}
\xi_T = \xi_g \left[ 1 + \beta \left( \frac{T}{300\,\mathrm{K}} - 1 \right) \right],
\end{equation}
allowing partial delocalization at elevated temperatures. The influence of an external magnetic field $B$ is captured through a phenomenological shrinkage of the localization length,
\begin{equation}
\xi(B) = \frac{\xi_T}{\sqrt{1 + (B/B_0)^2}},
\end{equation}
where $B_0$ sets the characteristic magnetic field scale~\cite{kramer1993localization,bassler1993charge}.

The phenomenological parameters used to describe the localization length and effective mass ($\alpha$, $\beta$, and $B_0$) were selected to maintain consistency with known experimental transport gaps and mobility edge behavior in disordered graphene \cite{lherbier2012transport}. Specifically, the temperature scaling factor $\alpha$ accounts for the thermal broadening of the wave function overlap, a characteristic feature of phonon-assisted hopping in 2D systems \cite{moazzami2024ultimate,kang2020phonon}. Similarly, the magnetic parameter $B_0$ reflects the characteristic field strength at which the magnetic length becomes comparable to the spatial extent of the localized states, thereby suppressing the hopping probability \cite{mucciolo}. While these values represent a simplified mapping of the complex many-body interactions, sensitivity tests conducted within our framework indicate that the qualitative trends in transmittance and current-voltage characteristics remain robust across a physically reasonable range of these constants.

\subsection{Kinetic Monte Carlo simulation}

Carrier dynamics are simulated using a continuous-time kinetic Monte Carlo algorithm~\cite{bortz1975new,andersen2019practical}. Carriers are injected stochastically from sites near the source boundary and evolve through successive hopping events selected according to their transition rates. The waiting time between events is sampled from
\begin{equation}
\Delta t = -\frac{\ln \eta}{\sum_j \Gamma_{i \rightarrow j}},
\label{eq:deltat}
\end{equation}
where $\eta \in (0,1)$ is a uniformly distributed random number. Each trajectory terminates when the carrier reaches the drain or when a maximum simulation time is exceeded. Statistical convergence is achieved by averaging over a large ensemble of independent trajectories.

\subsection{Effective transport observables}

Transport efficiency is quantified by defining an effective transmitance
\begin{equation}
T(V) = \frac{N_{\mathrm{pass}}}{N_{\mathrm{tot}}},
\end{equation}
where $N_{\mathrm{pass}}$ is the number of trajectories reaching the drain. An effective conductance is obtained using a Landauer-inspired scaling relation~\cite{landauer1957spatial,datta1997electronic},
\begin{equation}
G(V) = G_0 \, T(V), \qquad G_0 = \frac{2e^2}{h},
\end{equation}
and the steady-state current follows as
\begin{equation}
I(V) = G(V) \, V.
\end{equation}

Here, the Landauer formula is not interpreted as a quantum transmission formalism, but rather as a convenient normalization that maps the fraction of successful trajectories onto an effective conductance scale, enabling comparison across different transport regimes.

It is important to clarify the physical interpretation of the transmittance $T(V)$ in the context of our hopping model. While the Landauer-Büttiker formalism is strictly derived for coherent quantum transport \cite{datta1997electronic}, we employ it here as a convenient normalization to map the stochastic results onto a conductance scale. Our framework $T(V)$ does not represent the quantum-mechanical transmission probability for a single wave function. Instead, it serves as an effective measure of the successful carrier trajectories across the disordered landscape \cite{filoche2017localization,piccardo2017localization,li2017localization}. In this diffusive-to-localized limit, the current is governed by the probability of establishing a percolative path, where the "effective transmittance" reflects the ratio between the actual hopping-mediated current and the theoretical maximum current of a perfectly ballistic channel \cite{nourbakhsh2018charge,feng2023direct}. This mapping allows for a direct comparison between the impacts of external fields and structural disorder on a common dimensionless scale.

\subsection{Electronic structure and effective gap determination}

When required, the effective transport gap entering the hopping model is determined independently from the atomic structure using a one-orbital, distance-dependent tight-binding Hamiltonian~\cite{slater1954simplified,setyawan2010high}. The hopping integrals decay exponentially with interatomic separation, and the electronic band structure is evaluated along high-symmetry paths in the Brillouin zone. The resulting energy separation between the valence and conduction manifolds is interpreted as an effective localization scale and incorporated into the transport model only through the parameters $m_{\mathrm{eff}}$ and $\xi$, preserving a clear separation between electronic structure characterization and non-equilibrium transport simulation.

Within the methodological framework adopted here, electronic transport in graphene is treated in a regime where phase coherence is not preserved over macroscopic length scales. By considering temperatures starting from room temperature ($\approx$ 300 K), electron-phonon interactions become relevant and effectively suppress long-range quantum coherence ~\cite{tikhonenko2008weak, park2014electron}, precluding purely ballistic transport. As a result, charge propagation is described in terms of diffusive dynamics, which can be consistently approximated by a hopping-based formalism ~\cite{liang2012impurity}. This approach allows us to isolate disorder-induced localization effects while remaining physically consistent with experimentally accessible temperature regimes. Consequently, pristine graphene is not modeled as an ideal ballistic Dirac system, but rather as a reference limit within a broader diffusive transport framework, against which deviations driven by disorder and lattice connectivity can be systematically analyzed.

While quasi-ballistic effects may persist in micron-scale samples, the present work deliberately targets regimes in which disorder, thermal activation, and finite coherence lengths render transport effectively diffusive.

\section{Results and discussion}

We next define a graphene-based reference configuration that serves as the quantitative baseline for all subsequent perturbative studies (defects, strain, temperature, and magnetic field). Importantly, in the present work we do not treat the bandgap as a manually adjustable parameter. Instead, the gap is computed self-consistently from the atomic structure whenever the lattice is modified---specifically, upon the introduction of vacancies under externally applied strain. The resulting tight-binding (TB) gap is then injected into the hopping framework exclusively through the localization-controlled transport parameters, as detailed in the Methodology ~\cite{setyawan2010high}. This strategy ensures that changes in electronic structure arise from physically motivated structural perturbations, rather than from ad hoc tuning, and enables a direct link between atomic-scale modifications, wavefunction localization, and macroscopic transport observables.

\begin{figure}[b!]
\centering
\includegraphics[width=1.0\linewidth]{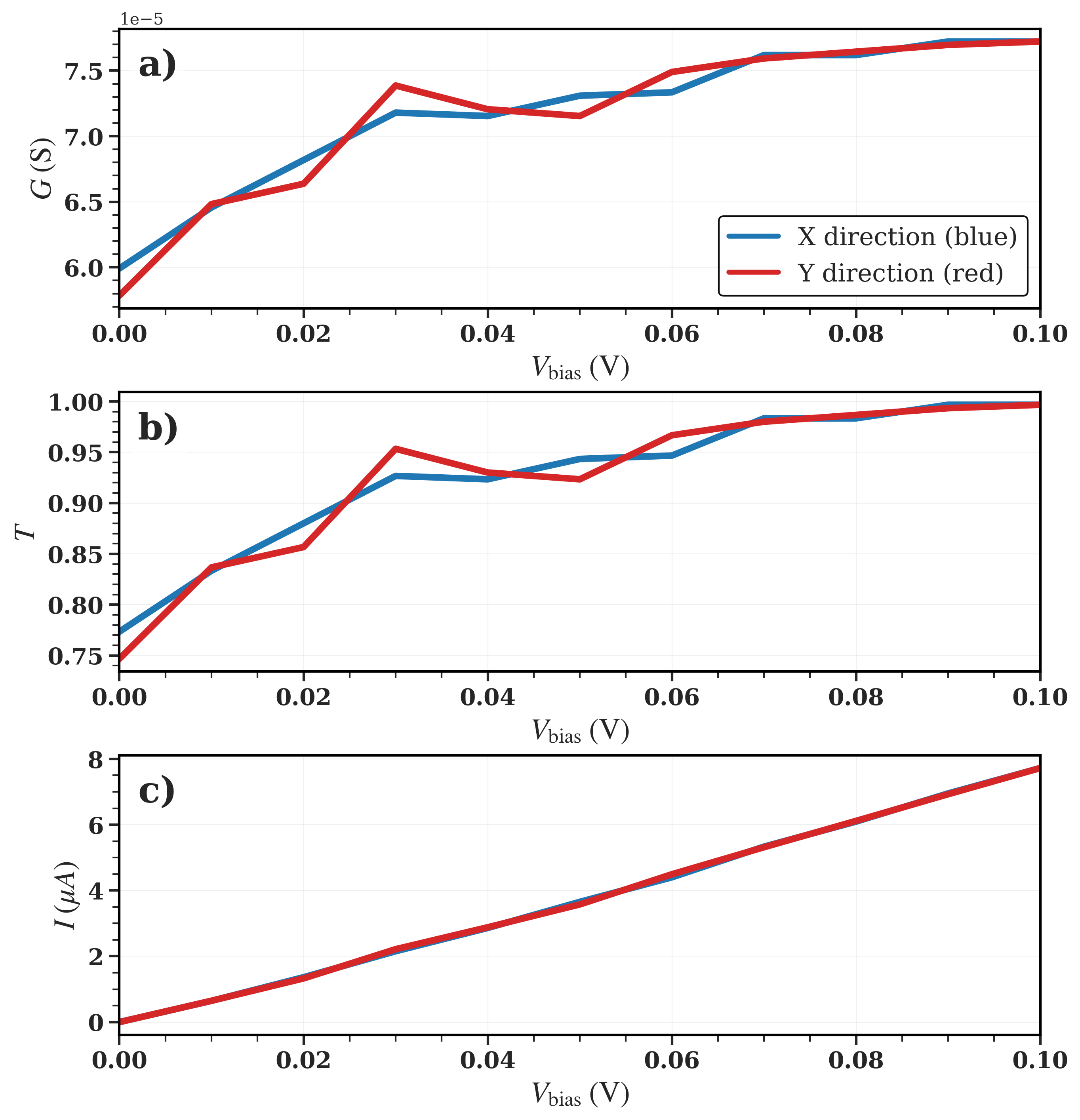}
\caption{Bias-dependent transport characteristics of the graphene reference system at zero magnetic field ($B=0$~T). The top panel shows the conductance $G(V)$, the middle panel the effective transmitance $T(V)$, and the bottom panel the resulting current--voltage characteristic $I(V)$, all plotted as functions of the applied bias $V$. The monotonic increase and subsequent saturation of $T(V)$ and $G(V)$ reflect the progressive enhancement of forward percolation across the hopping network under an increasing electrostatic energy tilt, while the approximately linear $I(V)$ response serves as a benchmark reference for subsequent analyses of localization effects induced by temperature, magnetic field, strain, and defect formation.}
\label{fig:fig1}
\end{figure}

The system consists of a $4\times7\times1$ graphene supercell (112 atoms), and the parameters are chosen to reproduce the characteristic low-bias, nearly ohmic response of pristine graphene.
Moreover, the replication strategy adopted in this work is not arbitrary but follows directly from the transport dynamics defined by Eq.~\ref{eq:deltat}. When different system lengths are imposed along the $X$ and $Y$ directions, an artificial separation of transport time scales is introduced. As a result, particle trajectories are forced to evolve differently along orthogonal directions within the same temporal window, leading to an anisotropic transport regime that does not reflect intrinsic structural properties but rather a geometric artifact of the simulation setup.

For this reason, in order to investigate isotropic and anisotropic transport effects in a controlled and physically meaningful manner, the system lengths along the $X$ and $Y$ directions must be identical, or at least comparable. This constraint ensures that any anisotropy observed in the transport properties originates exclusively from the underlying structure or microscopic interactions of the system, rather than from externally imposed geometric asymmetries. Consequently, the methodology employed here allows for an unambiguous attribution of anisotropic transport behavior to genuine material properties.

Figure~\ref{fig:fig1} summarizes the baseline transport response at $B=0$~T, reporting the conductance $G(V)$ (top panel), the effective transmitance $T(V)$ (middle panel), and the current $I(V)$ (bottom panel) as functions of the applied bias $V$. 
These three observables are connected through the Landauer-inspired mapping used in this study ~\cite{datta1997electronic} , $G(V)=G_0\,T(V)$ with $G_0=2e^2/h$, and $I(V)=G(V)\,V$. Accordingly, the bias dependence observed in the three panels provides a compact diagnostic of how the microscopic hopping dynamics evolve under the electrostatic energy tilt imposed by the bias.

The middle panel of Figure~\ref{fig:fig1} shows that the effective transmitance $T(V)$ increases monotonically from $\sim 0.75/0.77$ at the lowest biases to values approaching unity at higher bias within the explored range, isotropic along X and Y directions. In the present kinetic Monte Carlo formulation, $T(V)$ is defined as the fraction of injected trajectories that successfully traverse the device from the source boundary to the drain boundary. The observed increase of $T(V)$ with bias is therefore interpreted as a progressive enhancement of forward percolation across the hopping network: as $V$ increases, the linear potential drop lowers the energetic cost of forward hops relative to backward hops, increasing the likelihood that a carrier trajectory reaches the drain before becoming trapped or redirected. The approach to $T(V)\approx 1$ at a larger $V$ indicates that, for this reference graphene geometry and parameter set, essentially all injected carriers find a viable conductive pathway once the bias-induced energy tilt is sufficiently strong. This behavior is consistent with the expectation that pristine graphene, in the absence of strong localization sources, exhibits high transport efficiency under modest driving fields \cite{bolotin2014electronic, ElectronicTwoGraph}.

Consistent with the Landauer-inspired mapping, the top panel of Figure~\ref{fig:fig1} shows that $G(V)$ follows the same qualitative trend as $T(V)$: it increases with bias and then saturates as $T(V)$ approaches unity. The conductance plateau at larger $V$ reflects the saturation of transmission rather than a change in intrinsic channel count, as the model effectively treats the transport response as governed by the probability of successful traversal across a graph of localized states. In this sense, the plateau is a clear signature that the system has entered a bias regime where the microscopic hopping dynamics are no longer limited by connectivity or activation barriers within the simulated domain, but rather by the maximal transmission allowed by the given network and boundary definition. This provides an internally consistent baseline against which any subsequent suppression of $G(V)$ can be attributed to enhanced localization (e.g., due to defect-induced gap opening, strain-induced band-structure modification, or magnetic-field-induced shrinkage of the localization length).

The bottom panel of Figure~\ref{fig:fig1} reports the resulting $I(V)$ characteristics. The current grows approximately linearly across the studied bias range, with a slope that reflects the effective conductance. In practical terms, this behavior reproduces the hallmark of graphene-like ohmic response while still capturing the physically meaningful bias-dependent improvement in traversal probability intrinsic to a hopping network under an energy tilt \cite{ElectronicTwoGraph}. This is precisely the regime we seek as a reference point: it is sufficiently graphene-like to benchmark the model, yet sufficiently sensitive to localization physics to allow measurable deviations when perturbations are introduced.

The current levels obtained in our simulations, typically in the microampere range for biases up to 1 V, are consistent with experimental values reported for graphene field-effect transistors \citep{DiBartolomeo2016}.Although the bias scale differs due to distinct device geometries, the order of magnitude of the current agrees well with those measurements. Remarkably, the conductance of our pristine graphene reference ($\sim 80\,\mu$S) is within a factor of two of values measured in graphene/hBN/graphene tunnel transistors ($\sim 160\,\mu$S) \citep{greenaway2018tunnel}, despite the different material systems and transport regimes. This agreement supports the physical relevance of the parameters used in our hopping framework.

The choice of simulation size is similarly motivated by the need to balance physical representativeness with computational tractability for systematic parameter sweeps. We employ a $7\times 4\times 1$ supercell containing 112 atoms, which we verified to be sufficient to yield stable and reproducible $G(V)$, $T(V)$, and $I(V)$ responses in the low-bias regime without incurring excessive simulation time. 

We also check that an electrode width equal to 5~\AA, is sufficient to reproduce the ohmic behavior of graphene. Because the hopping probability depends exponentially on distance through $\exp(-2r/\xi)$, the bare localization length $\xi_0$ effectively sets the wavefunction extent relative to the device dimensions and thus controls the degree of network-wide connectivity. For substantially larger domains, $\xi_0$ renormalization may be required to preserve comparable transmission statistics and avoid artificial size-induced suppression of transport; however, for the comparative analyses pursued here, focused on relative changes induced by temperature $T$, magnetic field $B$, strain, and defect formation, the 112-atom configuration provides an efficient and reliable baseline.

In the remainder of this work, we exploit this baseline to interrogate how current is modulated by mechanisms that directly influence wavefunction localization and thus the hopping rates. Temperature and magnetic field are varied explicitly to probe, respectively, thermal activation and magnetic shrinkage of the effective localization length. Structural perturbations are introduced through vacancies, and externally applied strain; in these cases, the electronic bandgap is recalculated from the perturbed atomic structure using the same TB formalism, and the computed gap is then propagated into the hopping model via the localization and effective-mass renormalizations described previously. Throughout, this workflow maintains a direct, physically grounded causality chain: atomic-scale perturbation $\rightarrow$ TB electronic structure (gap) $\rightarrow$ modified localization length and hopping rates $\rightarrow$ changes in $T(V)$, $G(V)$, and ultimately $I(V)$, enabling a unified and interpretable analysis of localization-driven transport across graphene-derived systems.

\begin{figure}[b!]
\centering
\includegraphics[width=\linewidth]{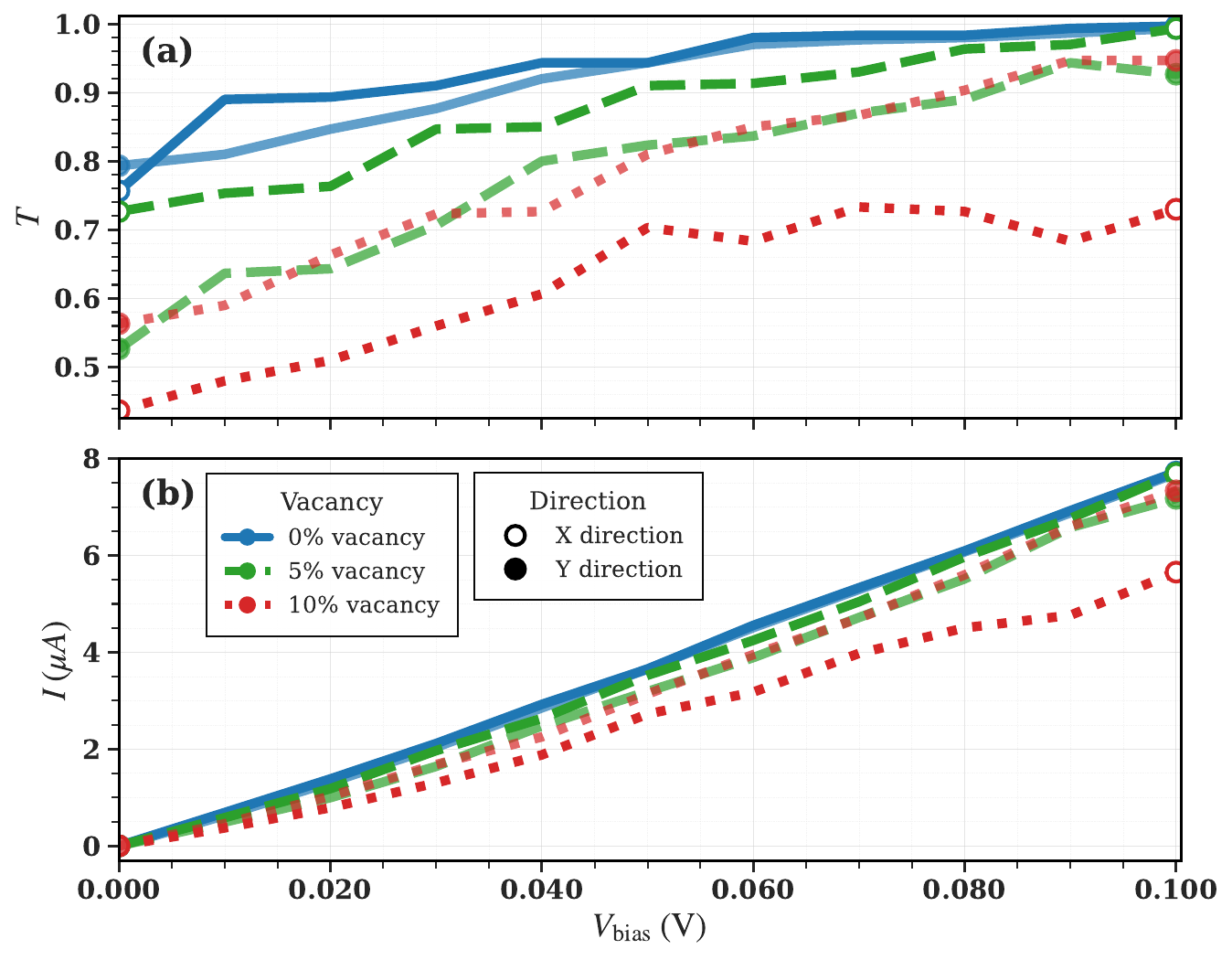}
\caption{Bias-dependent transport response of graphene under randomly distributed vacancies. Panel (a) shows the effective transmittance, $T(V)$ and panel (b) the corresponding current--voltage characteristics $I(V)$ for vacancy concentrations of 0\%, 5\%, and 10\%. Results are resolved along the $X$ and $Y$ transport directions, as indicated by open and filled symbols, respectively. Vacancies are introduced at random lattice sites, while all other parameters (temperature $T=300$ K, magnetic field $B=0$~T, electrostatic bias profile, and hopping model parameters) are kept identical to the reference system. Increasing vacancy concentration progressively suppresses transmittance and current, reflecting reduced lattice connectivity and enhanced localization. At higher disorder levels, statistically induced directional anisotropy emerges despite the absence of any externally imposed symmetry breaking, highlighting the sensitivity of hopping transport to the microscopic topology of the percolation network.}
\label{fig:fig2}
\end{figure}

From this point onward, we focus exclusively on the effective transmittance $T(V)$ as the primary microscopic transport observable. As demonstrated in Figure~\ref{fig:fig1}, the conductance $G(V)$ follows exactly the same bias dependence as $T(V)$ through the Landauer-inspired relation $G(V)=G_0 T(V)$. Consequently, reporting both quantities would be redundant and would not add additional physical insight. The transmittance, defined directly from the fraction of successful carrier trajectories, provides a more transparent connection between network connectivity, localization, and transport efficiency within the hopping framework.

Figure~\ref{fig:fig2} examines the impact of structural disorder introduced by randomly distributed vacancies on charge transport in graphene. The figure reports the bias-dependent effective transmitance $T(V)$ [panel \ref{fig:fig2}-a)] and the corresponding current--voltage characteristics $I(V)$ [panel \ref{fig:fig2}-b)], resolved along the $X$ and $Y$ transport directions. Vacancy concentrations of 0\%, 5\%, and 10\% are considered, with vacancy positions chosen randomly to avoid any artificial symmetry or directional bias. All other parameters, including temperature ($T=300$~K), magnetic field ($B=0$~T), electrostatic bias profile, and hopping model parameters, are identical across all cases.

In the absence of vacancies (0\%), the system exhibits high transmittance across the entire bias range, with $T(V)$ rapidly approaching unity for both transport directions. This behavior reflects the high connectivity of the pristine graphene lattice, where multiple percolation pathways are available and the bias-induced energy tilt efficiently promotes forward hopping. The near overlap between the $X$- and $Y$-direction curves confirms the expected isotropy of ideal graphene within the present framework and validates the reference configuration established in Figure~\ref{fig:fig1}.

Introducing a moderate vacancy concentration of 5\% leads to a noticeable suppression of transmittance. While $T(V)$ still increasing monotonically with bias, its saturation value is reduced relative to the pristine case, indicating that a fraction of carrier trajectories becomes trapped or redirected due to the removal of lattice sites. Importantly, transport remains largely isotropic: the differences between the $X$ and $Y$ directions are minor, suggesting that at this level of disorder the graphene network retains sufficient redundancy to compensate for randomly missing sites. This regime corresponds to a weakened but still percolating hopping network, in which localization effects are enhanced but do not yet dominate transport.

A key feature emerging at 10\% vacancy concentration is the appearance of a clear directional asymmetry between transport along $X$ and $Y$. While vacancies are introduced randomly, their statistical realization within a finite simulation domain can lead to anisotropic connectivity, favoring one transport direction over the other. Within the hopping framework, this manifests as direction-dependent transmittance and current, reflecting the sensitivity of localized transport to the detailed topology of the percolation network rather than to long-range crystalline symmetry. This anisotropy is not imposed externally but rather emerges naturally from the interplay between disorder and finite-size effects, highlighting the microscopic resolution of the approach.

The current--voltage characteristics shown in panel \ref{fig:fig2}-b) mirror these trends. For pristine graphene, the $I(V)$ curves are nearly linear and isotropic, consistent with graphene-like ohmic behavior. As vacancies are introduced, the slope of the $I(V)$ curves decreases, directly reflecting the reduction in transmittance. At 10\% vacancy concentration, the current is substantially suppressed and exhibits clear directional dependence, signaling a transition toward a transport regime dominated by localization and limited percolation rather than by uniform network connectivity.

\begin{figure*}[!t] \centering \includegraphics[width=\linewidth]{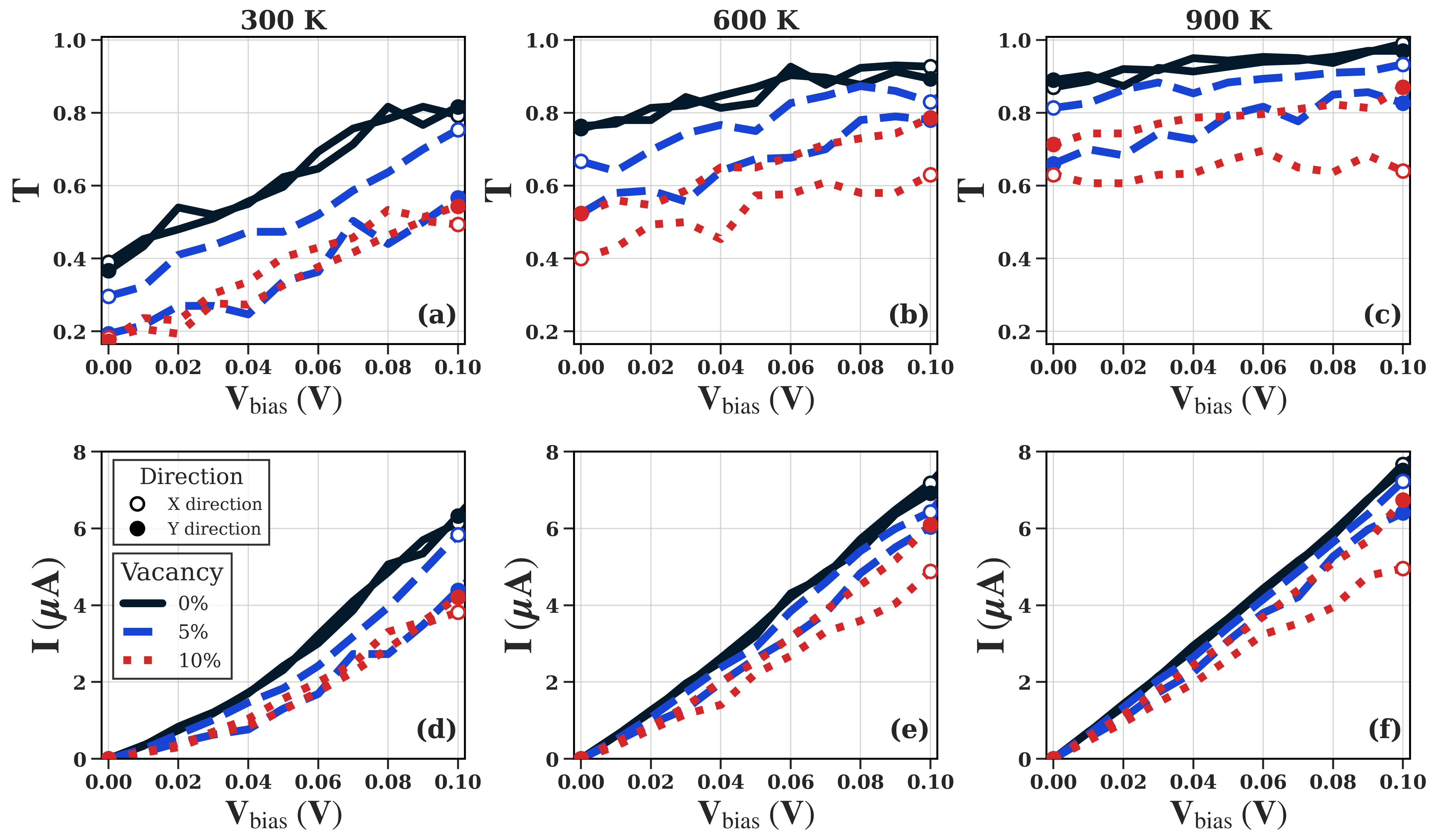}
\caption{Temperature-dependent transport response of graphene with randomly distributed vacancies. Top panels show the effective transmitance $T(V)$ and bottom panels show the corresponding current--voltage characteristics $I(V)$ for vacancy concentrations of 0\%, 5\%, and 10\%, resolved along the $X$ and $Y$ transport directions. Results are presented for temperatures of 300~K (left), 600~K (center), and 900~K (right), while all other parameters, including atomic structure, electrostatic bias profile, and magnetic field ($B=0$~T), are kept fixed. Increasing temperature systematically enhances transmittance and current by promoting thermally assisted hopping and partial delocalization, with the strongest relative effect observed at intermediate and high vacancy concentrations. The incomplete recovery of transport at high disorder highlights the dominant role of lattice connectivity in limiting charge transport, even under strong thermal activation.}
\label{fig:fig3}
\end{figure*}

Figure~\ref{fig:fig3} extends the vacancy-controlled analysis of Fig.~\ref{fig:fig2} by introducing temperature as an independent tuning knob for localization and hopping kinetics. The figure reports the effective transmittance $T(V)$ (top row) and the corresponding current–voltage characteristics $I(V)$ (bottom row) for vacancy concentrations of 0\%, 5\%, and 10\%, resolved along the $X$ and $Y$ directions, at $T = 300$~K, 600~K, and 900~K. This temperature sweep provides a direct and physically transparent probe of thermally assisted transport in a structurally disordered graphene network, allowing connectivity-limited suppression (set primarily by vacancy concentration) to be disentangled from activation- and delocalization-driven enhancement controlled by temperature.

Across all vacancy concentrations, increasing temperature systematically enhances both transmittance and current, with the effect being most pronounced at low to intermediate bias. In the pristine lattice (0\% vacancy), the network remains highly connected, and transport is already efficient at room temperature. Nevertheless, the upward shift of $T(V)$ with increasing $T$ indicates that thermal broadening and enhanced hopping activity further facilitate forward percolation. In this regime, temperature primarily accelerates transport dynamics rather than rescuing an impaired network, leading to only moderate relative gains and a response that remains nearly isotropic.

At 10\% vacancy, the system enters a regime dominated by fragmentation of conductive pathways and strong localization. At 300~K, $T(V)$ is strongly suppressed, and the current remains low, indicating that a large fraction of trajectories fail to reach the drain even under increasing bias. Upon heating to 600~K and 900~K, both $T(V)$ and $I(V)$ increase significantly, demonstrating that temperature can partially re-enable transport by allowing carriers to escape local traps and explore alternative routes through the disordered network. Crucially, however, this recovery remains incomplete: even at the highest temperature considered, the 10\% vacancy curves lie well below the pristine case. This establishes a clear hierarchy of control parameters in the present framework—temperature can accelerate and assist hopping dynamics, but it cannot reconstruct missing topological connectivity.
The enhancement of transport with temperature, particularly pronounced at intermediate disorder levels, is a hallmark of variable-range hopping and thermally activated conduction, as extensively documented in disordered graphene systems \cite{shklovskii1984variable, qiu2013hopping}. Recent experimental studies on defected graphene have also reported similar thermally activated behavior~\cite{PhysRevB.80.153404}.

A second key outcome concerns the evolution of directional anisotropy with temperature. In the pristine system, the $X$ and $Y$ responses remain closely overlapped across all temperatures, reflecting the intrinsic isotropy of graphene when connectivity is intact. With increasing vacancy concentration, finite-size disorder realizations generate anisotropic percolation backbones, leading to measurable directional differences in both $T(V)$ and $I(V)$. Notably, increasing temperature does not eliminate these anisotropies but progressively reduces their role as the dominant transport-limiting factor. Thermally assisted hopping broadens the ensemble of accessible trajectories, enabling charge carriers to bypass locally unfavorable paths without collapsing the underlying structural asymmetry imposed by disorder.
54
From a high-level perspective, Fig.~\ref{fig:fig3} establishes temperature as a powerful kinetic lever that modulates transport primarily through trajectory-level accessibility rather than through any modification of the underlying structural topology. The combined vacancy–temperature trends reveal a clear crossover: in highly connected networks (0\% vacancy), temperature yields only incremental improvements; in moderately disordered networks (5\% vacancy), thermal activation strongly enhances percolation and restores a near--graphene--like response; and in strongly disrupted networks (10\% vacancy), temperature produces only partial recovery, exposing a connectivity-imposed ceiling on transport. This redistribution of transport pathways, rather than their homogenization, shifts the dominant limitation from directional bottlenecks to global topological constraints.

The resulting interplay between topology (vacancies) and kinetics (temperature) provides a physically transparent framework to rationalize how realistic graphene devices evolve from near-ohmic conduction to activated, localization-dominated transport as disorder and thermal effects increase.

\begin{figure*}[!t]
\centering
\includegraphics[width=1.0\linewidth]{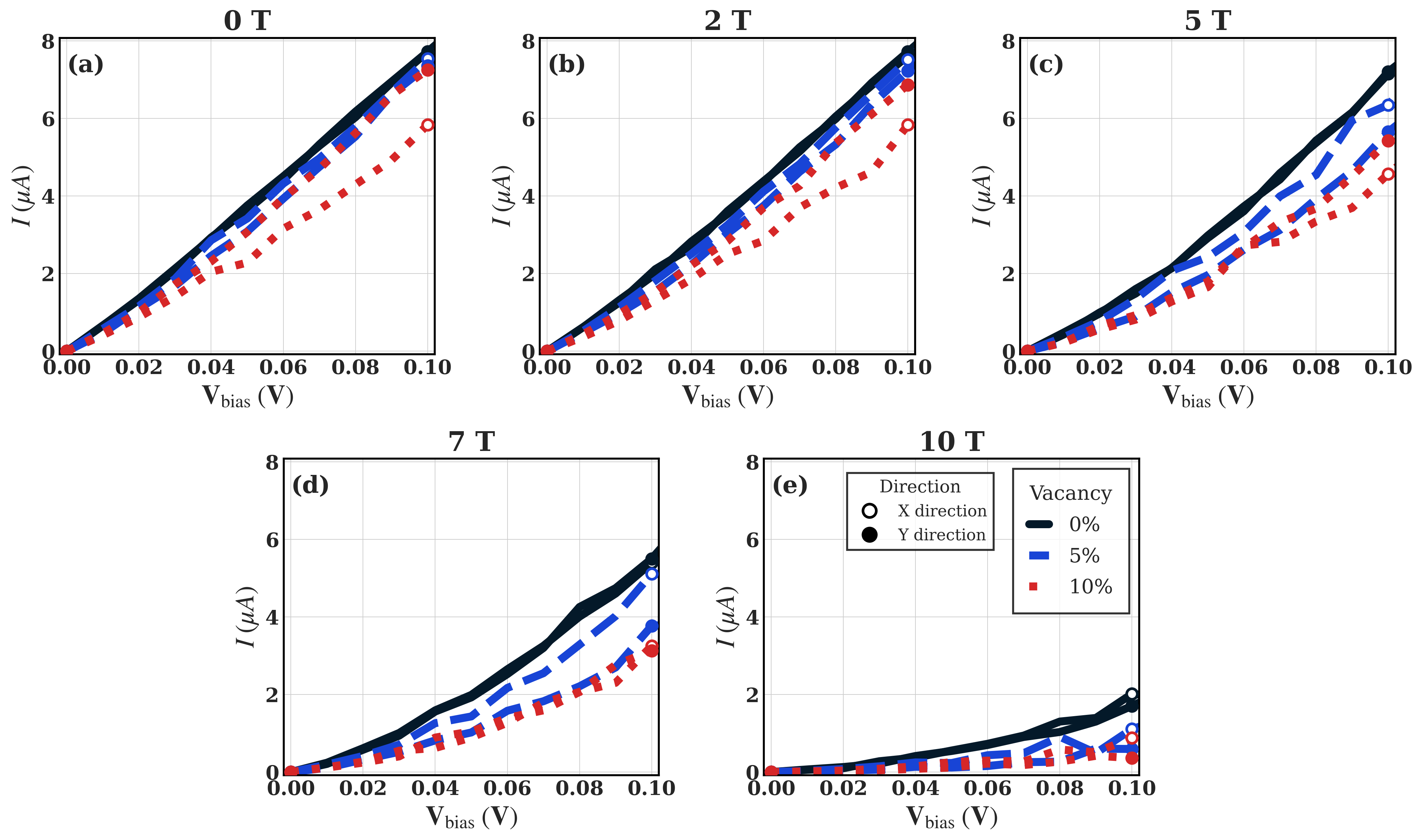}
\caption{Magnetic-field-dependent current--voltage characteristics of graphene with randomly distributed vacancies. The current $I(V)$ is shown for vacancy concentrations of 0\%, 5\%, and 10\%, resolved along the $X$ and $Y$ transport directions, under external magnetic fields $B=0, 2, 5, 7,$ and $10$~T. While pristine graphene retains a quasi-linear response at low magnetic fields, increasing $B$ leads to a pronounced suppression of current across the entire bias range. The effect is strongly enhanced by vacancies, which reduce network redundancy and magnify the impact of magnetic-field-induced localization. Directional asymmetries emerging at high field and disorder levels highlight the sensitivity of hopping transport to the microscopic topology of the percolation network.}
\label{fig:fig4}
\end{figure*}

 Figures~\ref{fig:fig4sup} and~\ref{fig:fig4} quantify how an external magnetic field in average directions progressively suppresses hopping transport in graphene networks as structural connectivity is degraded by randomly distributed vacancies. In this section we deliberately split the presentation across the Supplementary Material and the main text: the effective transmittance $T(V)$ is reported in Figure~\ref{fig:fig4sup} (Supplementary) because it mirrors the conductance trends and serves primarily as a microscopic diagnostic of traversal statistics, whereas the experimentally closest observable—the current $I(V)$—is emphasized in the main text in Figure~\ref{fig:fig4}. In both figures, transport is resolved along the $X$ and $Y$ directions, and vacancy concentrations of 0\%, 5\%, and 10\% are compared under magnetic fields $B=0,2,5,7,$ and $10$~T.

Figure~\ref{fig:fig4sup} reveals a systematic, field-driven reduction of $T(V)$ across all vacancy concentrations, consistent with magnetically induced localization entering the hopping rates through the localization length $\xi(B)$ \cite{raikh1993single}. At $B=0$~T the pristine lattice (0\% vacancy) exhibits high transmittance that increases with bias and approaches unity, reflecting strong network redundancy and efficient forward percolation under the electrostatic energy tilt. As $B$ increases to 2~T and 5~T, $T(V)$ is suppressed but remains appreciable, indicating that the network still supports connected traversal pathways; the bias dependence remains monotonic, showing that the electric-field tilt can partially compensate for the reduced overlap by preferentially promoting forward hops. At 7~T, the suppression becomes severe, and the accessible percolation backbone effectively collapses into a narrower set of viable paths, with $T(V)$ remaining substantially below unity even at the highest bias. At 10~T, transmittance becomes strongly quenched for all vacancy concentrations, and only a small fraction of trajectories reach the drain within the explored bias window—an unambiguous signature that magnetic confinement has reduced wavefunction overlap to the point that connectivity alone cannot sustain traversal.

Crucially, the magnetic-field response is strongly amplified by disorder. While the pristine network retains relatively higher $T(V)$ at intermediate fields, vacancy-containing networks show a markedly steeper degradation with $B$. This reflects a multiplicative interplay between topology and localization: vacancies remove nodes and edges from the transport graph (reducing the number of alternative paths), while the magnetic field reduces the effective hopping range (shrinking $\xi$), thereby penalizing the remaining longer or more tortuous routes that would otherwise bypass missing sites. The combined effect is not merely additive; rather, $B$ selectively eliminates the marginal pathways that are most essential for maintaining percolation in a fragmented lattice, making magnetolocalization disproportionately effective at higher vacancy concentrations. The observed crossover from a linear to a strongly nonlinear current-voltage response with increasing disorder is in good agreement with experimental studies of defected graphene \cite{SOTOUDEH2018828} and aligns with theoretical models of variable-range hopping and localization in disordered systems \cite{shklovskii1984variable}, \cite{PhysRevLett.96.036801}.

Directional differences between $X$ and $Y$ remain small for 0\% vacancy across fields, consistent with the near-isotropy of pristine graphene. With vacancies, modest $X/Y$ asymmetries become increasingly visible as $B$ they grow, particularly at high fields where transport is forced to rely on a sparse percolation backbone. In this regime, finite-size statistical fluctuations in the random vacancy realization translate into direction-dependent bottlenecks. Importantly, this anisotropy is emergent: it is not imposed by any anisotropic parameter choice but arises from the microscopic topology of the remaining connected subgraph under strong localization constraints.

The current characteristics in Figure~\ref{fig:fig4} translate these trajectory-level trends into an experimentally interpretable magnetotransport signature. At $B=0$~T, the $I(V)$ curves are quasi-linear for 0\% vacancy and progressively reduced for 5\% and 10\% vacancy, consistent with the disorder-driven suppression already established in Figure~\ref{fig:fig2}. Increasing $B$ produces a clear reduction in the slope of $I(V)$ for all vacancy levels, i.e., a strong negative magnetoconductance within the present hopping framework. At low-to-moderate fields (2~T and 5~T), the suppression is substantial, yet the current remains finite and retains a broadly increasing trend with bias, indicating that field-driven localization has not completely destroyed forward percolation but has reduced its efficiency. At 7~T, the current is dramatically reduced across the entire bias range and begins to show stronger curvature, reflecting the transition from an effectively well-connected regime to one in which current is limited by rare, bias-assisted traversals through a strongly constrained network. At 10~T, the current is nearly extinguished for all vacancy concentrations within the explored voltages, consistent with the near-vanishing $T(V)$ observed in Figure~\ref{fig:fig4sup}.

Two mechanistic points stand out. First, the field dependence is highly nonlinear: modest increases in $B$ already large fields produce disproportionately large drops in $I(V)$. This behavior is expected when hopping probabilities depend exponentially on $\xi(B)$: once $\xi$ falls below a topology-dependent threshold set by typical inter-site separations and the availability of alternative routes; the network effectively crosses into a magnetically localized regime where current is controlled by extremely rare successful trajectories. Second, disorder increases magnetic sensitivity. The strong negative magnetoconductance observed in Fig.~4, with current suppression becoming disproportionately large at high fields, is consistent with magnetoresistance measurements in disordered graphene~\cite{gorbachev2007weak, chen2010magnetoresistance}. Theoretical models of magnetic-field-induced localization in two-dimensional systems predict exactly this type of behavior when the magnetic length becomes comparable to the localization length~\cite{abrahams1979scaling, kramer1993localization}. The separation between 0\%, 5\%, and 10\% vacancy becomes more consequential as $B$ grows, because vacancies remove the redundancy that would otherwise allow carriers to compensate for reduced hopping range. In other words, a pristine lattice can tolerate a reduction in $\xi$ by rerouting through multiple equivalent pathways, whereas a vacancy-fragmented lattice cannot, so magnetic-field-induced shrinkage of $\xi$ more readily produces a connectivity crisis.

Finally, the $X/Y$ comparison in Figure~\ref{fig:fig4} reinforces the interpretation that anisotropy is topology-driven. At low field, currents along $X$ and $Y$ remain close, particularly in the pristine case. Under strong fields and higher vacancy concentrations, direction-specific deviations become more apparent, reflecting the fact that the dominant current is carried by a small number of critical paths whose existence and quality depend on the particular defect realization and device orientation. This provides an important conceptual bridge to real devices: even when disorder is nominally isotropic on average, transport under strong localization conditions can become sample-specific and direction-sensitive because conduction is controlled by a sparse subset of percolation channels.

\begin{figure*}[!t]
\centering
\includegraphics[width=1.0\linewidth]{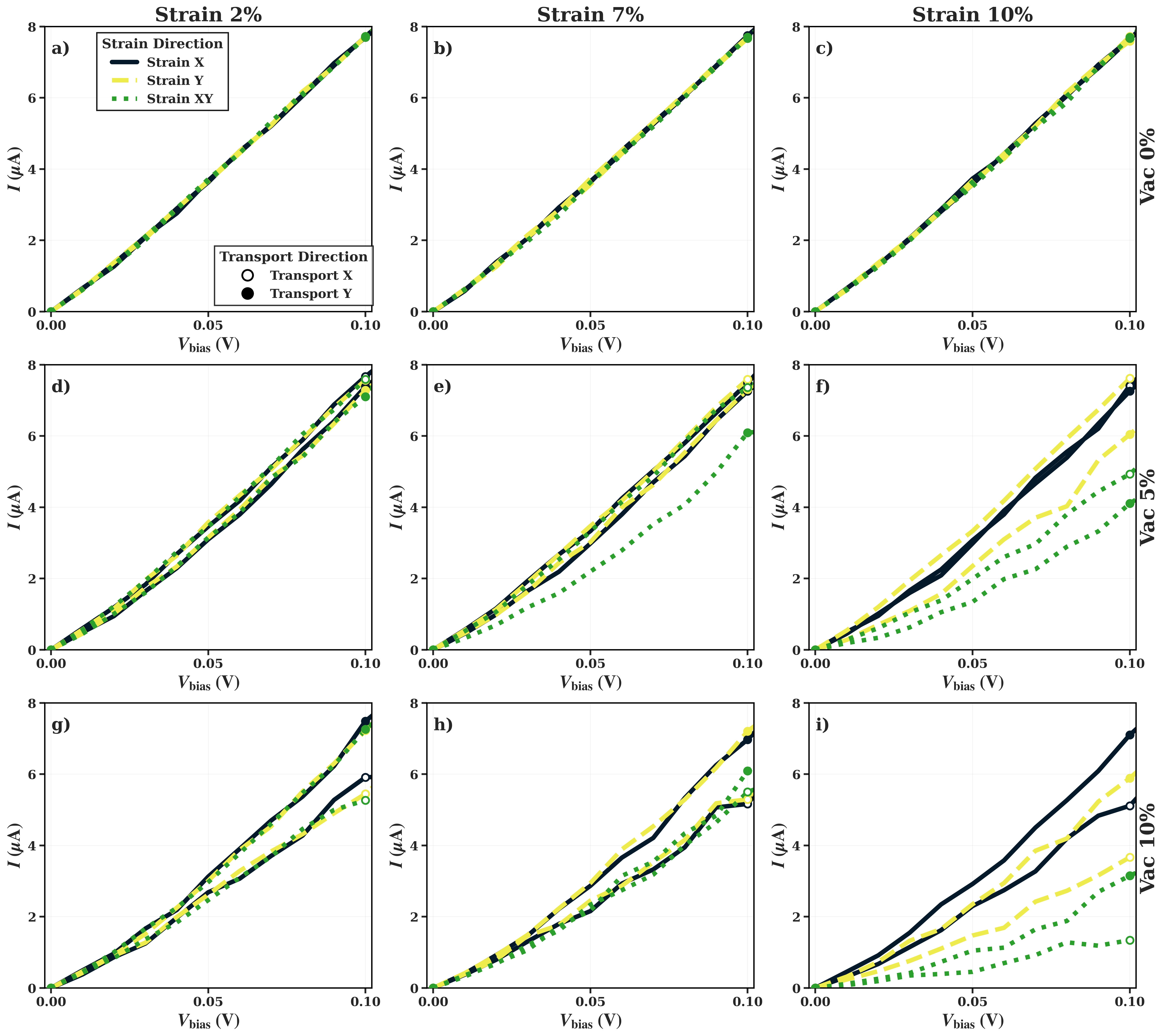}
\caption{Current--voltage characteristics of graphene under combined structural disorder and mechanical strain. The current $I(V)$ is shown for vacancy concentrations of 0\%, 5\%, and 10\%, strain magnitudes of 2\%, 7\%, and 10\%, and strain applied along the $X$, $Y$, or both in-plane directions ($XY$). Transport is resolved along the $X$ and $Y$ directions. While pristine graphene remains largely insensitive to strain within the studied range, vacancy-containing lattices exhibit a strong, orientation-dependent suppression of current as strain increases. This behavior reflects a nonlinear interplay between reduced lattice connectivity and strain-enhanced localization, which progressively shifts transport from a redundancy-dominated regime toward one controlled by a sparse, strain-sensitive percolation backbone.}
\label{fig:fig5}
\end{figure*}

Figures~\ref{fig:fig5sup} (Supplementary) and~\ref{fig:fig5} (main text) examine how mechanical strain reshapes hopping transport in graphene when structural disorder is simultaneously present. In direct analogy with the magnetic-field analysis (Figures~\ref{fig:fig4sup} and~\ref{fig:fig4}), we report the effective transmitance $T(V)$ as a microscopic diagnostic in the Supplementary Material, while emphasizing the current $I(V)$ in the manuscript as the experimentally closest observable. The figure set spans vacancy concentrations of 0\%, 5\%, and 10\% (rows), strain magnitudes of 2\%, 7\%, and 10\% (columns), and three strain orientations: uniaxial along $X$, uniaxial along $Y$, and biaxial ($XY$). Transport is additionally resolved along the $X$ and $Y$ directions, allowing us to disentangle (i) the direction in which the lattice is strained from (ii) the direction along which carriers traverse the device.

The observed anisotropy in transport properties, particularly at higher vacancy concentrations (e.g., 10\%), warrants a discussion on finite-size effects and statistical realizations. In our simulations, the specific spatial distribution of vacancies within the finite supercell can lead to the formation of preferential percolation paths or localized bottlenecks along a specific axis \cite{dietl2009disorder}. While such directional dependence might diminish in the macroscopic limit—where disorder averages out—it remains highly relevant for nanoscopic devices where the specific configuration of defects (the "fingerprint" of disorder) dictates the device's performance \cite{Rycerz2007}. To ensure the physical validity of these results, the reported trends reflect the ensemble behavior of stochastic trajectories, highlighting that in the hopping regime, the connectivity of the underlying network is as critical as the carrier density itself \cite{mucciolo}.

Within our modeling strategy, strain is not a purely geometric perturbation. It acts through two coupled microscopic channels: (i) it modifies inter-site distances and therefore the hopping network geometry, and (ii) for vacancy-containing structures, it alters the tight-binding electronic structure from which an effective transport gap is extracted and injected back into the hopping model via the localization length and effective-mass renormalizations. As a result, strain can either preserve transport (if connectivity remains high and the effective gap is small) or strongly suppress it (if strain enhances localization and/or amplifies defect-induced gap formation). This duality is precisely what Figures~\ref{fig:fig5sup} and~\ref{fig:fig5} resolve.

For 0\% vacancy (top row), both $T(V)$ and $I(V)$ remain high across all strain magnitudes and strain directions. Transmitance rapidly approaches unity with increasing bias, and the $I(V)$ curves remain nearly linear and almost indistinguishable across strain orientations. This indicates that, in the absence of vacancy-induced topology loss, the graphene network retains sufficient redundancy that modest-to-moderate strain (up to 10\%) does not qualitatively disrupt percolation within the present hopping description. In high-connectivity networks, the bias-induced energy tilt dominates transport, and strain-induced geometric distortions merely perturb local hop distances without producing a system-wide connectivity crisis. The near overlap between transport along $X$ and $Y$ further indicates that, for pristine lattices and the present device size, strain does not introduce a strong emergent anisotropy in transport efficiency.

At 5\% vacancy (middle row), the impact of strain becomes substantially more structured. While transport remains viable at low strain (2\%), differences between strain orientation and transport direction begin to emerge as strain increases to 7\% and 10\%. In this regime, vacancies already reduce the number of available conductive backbones, so strain can more effectively ``steer'' current by preferentially enhancing or degrading the overlap along specific directions. Uniaxial strain can elongate bonds and increase the effective hop distances along its principal axis, which directly suppresses hopping rates through the exponential distance factor. Simultaneously, strain can alter the defect-modified electronic structure and thereby shrink the localization length, further suppressing long hops that would otherwise bypass vacancy-induced bottlenecks. The net result is a strain-tunable competition between (i) field-assisted forward percolation and (ii) strain-enhanced localization, producing a noticeable separation among the curves at higher strain.

Biaxial ($XY$) strain is particularly informative because it simultaneously perturbs both in-plane directions. The data show that, when vacancies are present, biaxial strain tends to yield the strongest transport suppression (most clearly visible in $T(V)$ in Figure~\ref{fig:fig5sup} and in the reduced $I(V)$ slopes in Figure~\ref{fig:fig5}). This is consistent with the intuition that biaxial strain removes the possibility of ``escaping'' into a less-perturbed direction: if both axes are strained, the network-wide reduction in overlap is more uniform and detour pathways are penalized more strongly.

The observed suppression of transmittance under strain, particularly in the biaxial configuration with high vacancy concentration, can be understood through the modification of the electronic landscape. While pristine graphene requires substantial uniaxial strain to undergo a metal-insulator transition \cite{pereira2009tight}, the presence of structural vacancies breaks the sublattice symmetry and creates a disordered potential \cite{rajasekaran2016effect}. In this context, the applied strain acts as a second-order perturbation that further renormalizes the hopping integrals between localized sites. This synergy between mechanical deformation and point defects promotes the emergence of an effective mobility gap, where the spatial overlap of electronic states is hindered by both the increased interatomic distances and the strain-induced localization \cite{berdyugin2022out}. Our results indicate that biaxial strain is more effective at suppressing transport because it disrupts the percolative network isotropically, maximizing the impact of existing vacancy clusters on the total current.

The most striking behavior occurs at 10\% vacancy (bottom row), where the lattice is already close to a percolation-limited regime. Here, strain acts as a powerful amplifier of disorder. Even when transport remains measurable at 2\% strain, increasing strain to 7\% and especially 10\% produces a dramatic reduction in transmittance and current, with strong separation between strain directions and between transport directions. This reflects a genuine crossover: the current becomes controlled by a sparse set of critical paths that are extremely sensitive to small geometric changes and to strain-induced localization enhancement. In this regime, strain does not simply reduce mobility; it effectively restructures the percolation backbone by eliminating marginal links, converting what were previously viable traversal routes into disconnected or energetically inaccessible segments.

A central mechanistic conclusion is that the combined action of vacancies and strain is highly nonlinear. Vacancies alone reduce network redundancy, but transport can still proceed through alternative routes. Strain alone (in the pristine lattice) produces only mild changes. However, when both are present, strain can selectively suppress precisely the long-range or detour hops required to maintain percolation in a fragmented network. This synergy naturally produces the strongest deviations from the graphene--like response in the high-vacancy, high-strain limit, consistent with the pronounced suppression seen in both $T(V)$ and $I(V)$.

Because the data explicitly resolve both the strain direction (how the lattice is deformed) and the transport direction (how carriers traverse), Figures~\ref{fig:fig5sup} and~\ref{fig:fig5} highlight an important conceptual point: anisotropy is not an externally imposed model artifact but an emergent consequence of topology plus deformation. Uniaxial strain tends to produce stronger suppression when transport is forced to proceed along the strained axis, whereas transport along the transverse direction can remain comparatively more efficient if alternative pathways survive. This directional selectivity becomes increasingly pronounced as vacancy concentration grows, confirming that the system transitions from an isotropic, redundancy-dominated regime to an anisotropic, backbone-dominated regime where strain orientation and defect topology jointly determine the dominant conduction channels.

\begin{figure*}[!htb]
\centering
\includegraphics[width=1.0\linewidth]{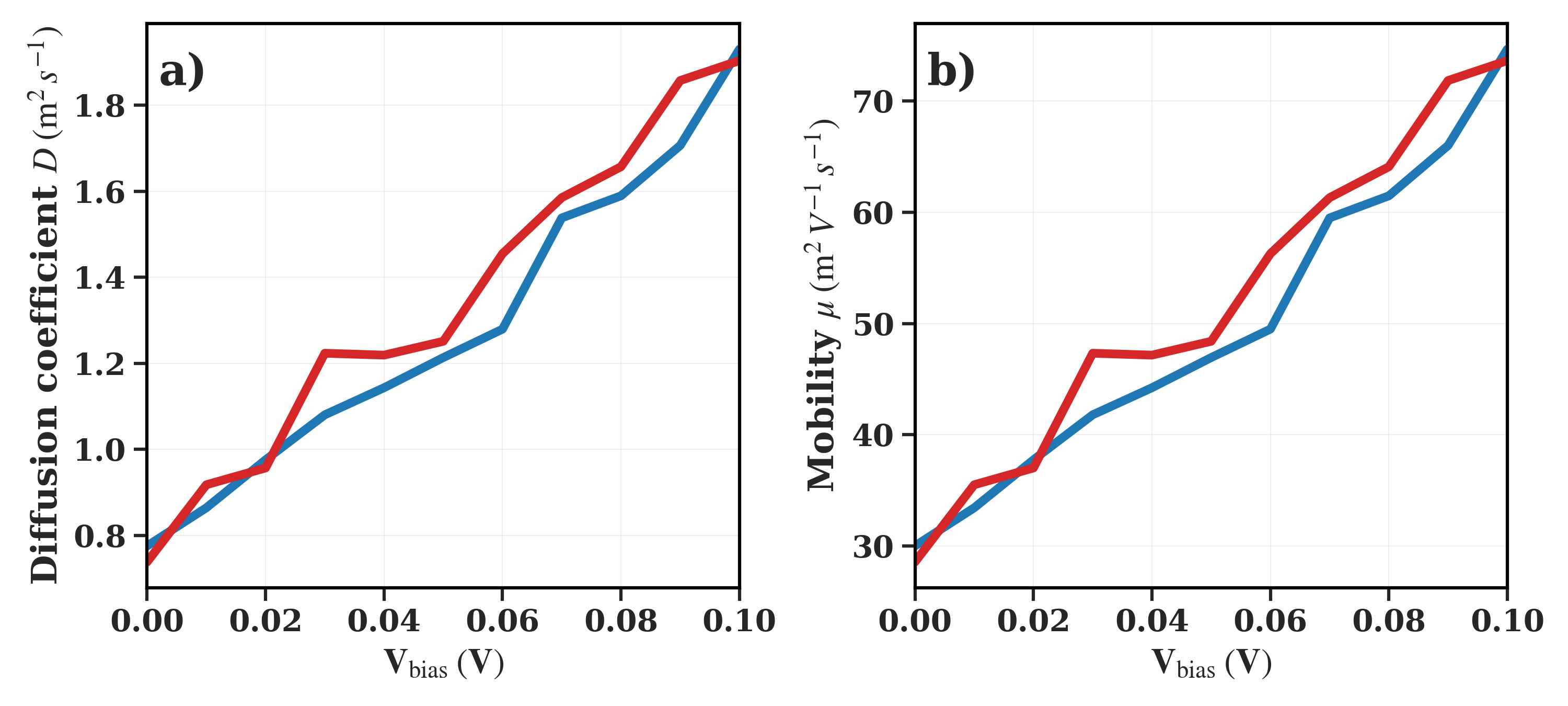}
\caption{Bias dependence of the effective diffusion coefficient $D$ (a) and the corresponding mobility $\mu$ (b) for pristine graphene at $T = 300$~K, zero magnetic field, and in the absence of vacancies. The diffusion coefficient is extracted from the kinetic Monte Carlo trajectories as $D = \langle (\Delta x)^2 \rangle / (2\,\tau)$, where $\Delta x$ denotes the carrier displacement and $\tau$ is the mean transit time. The mobility is obtained via the Einstein relation $\mu = eD/(k_B T)$. Here, $D$ is reported in units of m$^2$/s and $\mu$ in units of m$^2$/(V\,s), and both quantities should be interpreted as effective, diffusion-based transport parameters appropriate for incoherent, thermally activated hopping transport.}
\label{fig:Dmu}
\end{figure*}

Finally, the trajectory-resolved nature of the present framework allows the investigation of additional transport observables beyond transmittance and current. Figure~\ref{fig:Dmu} illustrates this capability by reporting the effective diffusion coefficient,
\begin{equation}
D = \frac{\langle (\Delta x)^2 \rangle}{2\,\tau},
\end{equation}
where $\Delta x$ denotes the carrier displacement and $\tau = \langle t \rangle$ is the mean transit time extracted from the kinetic Monte Carlo trajectories, together with the corresponding mobility,
\begin{equation}
\mu = \frac{e\,D}{k_B T},
\end{equation}
for pristine graphene at $T = 300$~K, zero magnetic field, and in the absence of vacancies \cite{baranovskii1996einstein}.

Both $D(V)$ and $\mu(V)$ exhibit smooth, monotonic increases with bias, reflecting the progressive enhancement of carrier drift and spatial exploration under the applied electric field. Importantly, these quantities are not imposed as phenomenological inputs but emerge directly from the underlying stochastic dynamics, reinforcing the internal consistency and physical transparency of the approach. The bias dependence further highlights the continuous crossover between diffusion-dominated and drift-assisted transport regimes, even in a structurally pristine lattice.

More generally, the results in Fig.~\ref{fig:Dmu} demonstrate that the proposed framework provides access to a broader hierarchy of transport descriptors derived from the same set of carrier trajectories. This enables a unified analysis of current, transmittance, diffusion, and mobility within a single modeling scheme and establishes the method as a versatile tool for exploring charge transport in two-dimensional carbon systems under varying bias, temperature, magnetic field, disorder, and lattice perturbations.

\section*{Conclusion}

In this summary, we presented a trajectory-resolved framework for charge transport in graphene and analogous two-dimensional carbon systems, intended to function beyond the idealized ballistic and fully coherent thresholds. The method uses an explicit atomic lattice representation and kinetic Monte Carlo hopping dynamics to clearly show how disorder, thermal activation, and external fields interact. Transport arises directly from stochastic carrier trajectories, facilitating the evaluation of current and effective transmittance without the necessity of phenomenological mobility or diffusion parameters.

We applied the framework to systematically quantify how bias voltage, temperature, magnetic field, mechanical strain, and vacancy concentration jointly control transport in graphene networks over practical ranges: $V_{\mathrm{bias}}=0$--$0.10$ V, $T=300$--$900$ K, $B=0$--$10$ T, in-plane strain of $2$--$10\%$ (uniaxial and biaxial), and vacancy concentrations of $0$--$10\%$. For pristine graphene, the response is almost ohmic over the bias window we examined. At 0.10 V, the current is about 7--8 $\mu$A, and the effective transmittance is about 0.98–1.00 (depending on the direction). The conductance in the same range is about $(5.8$--$7.8)\times 10^{-5}$ S. Vacancies gradually impede transport: at a 10\% vacancy rate, transmittance may decrease to approximately 0.45 to 0.75 (varying with bias and direction), accompanied by a significant reduction in current compared to the pristine condition.

The simulations show clear crossovers between regimes dominated by near-ohmic conduction, thermally assisted percolation, and localization. As the temperature rises, hopping kinetics speed up and other conducting pathways are encouraged. This partially recovers both $T$ and $I$ from $300$ K to $900$ K, but it doesn't fully compensate for the loss of connectivity at high defect densities, suggesting that the topology limits transport. External magnetic fields introduce an additional suppression that is strongest at high fields (up to $10$ T) and in less ordered networks. This makes it even harder for percolation to occur.

Beyond current--voltage characteristics, the trajectory-based formulation naturally provides access to a broader set of transport descriptors, including diffusion coefficients and effective mobilities derived directly from carrier displacements and transit times. In general, the method combines current, transmittance, diffusion, and mobility into a single modeling scheme. It also gives researchers a flexible way to study how charge moves through real two-dimensional carbon materials under the simultaneous influence of mechanical, thermal, and electromagnetic forces.



\pagebreak
\section{Supplentary Materials}

\begin{figure}[b!]
\centering
\includegraphics[width=1.0\linewidth]{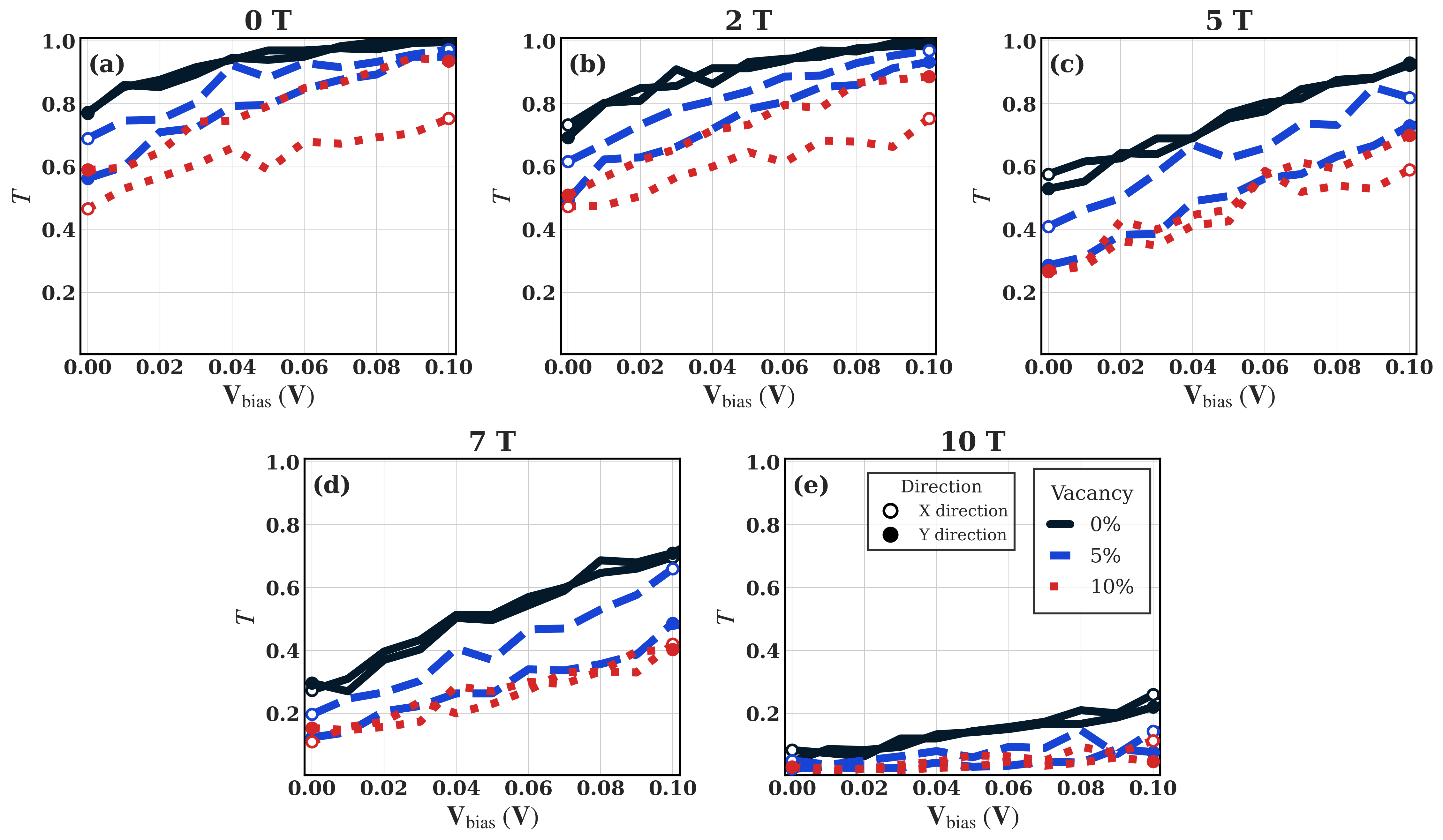}
\caption{Bias-dependent effective transmitance $T(V)$ of graphene under a perpendicular magnetic field for different vacancy concentrations. Results are shown for vacancy concentrations of 0\%, 5\%, and 10\%, with transport resolved along the $X$ and $Y$ directions, for magnetic field strengths $B=0, 2, 5, 7,$ and $10$~T. Vacancies are randomly distributed within the lattice, while temperature, electrostatic bias profile, and all hopping-model parameters are kept fixed. Increasing magnetic field systematically suppresses transmitance by reducing the effective localization length and thus the spatial overlap between localized states. The suppression is strongly amplified by structural disorder, reflecting the combined impact of reduced lattice connectivity and magnetically induced localization on percolative hopping transport.}
\label{fig:fig4sup}
\end{figure}

\begin{figure}[b!]
\centering
\includegraphics[width=1.0\linewidth]{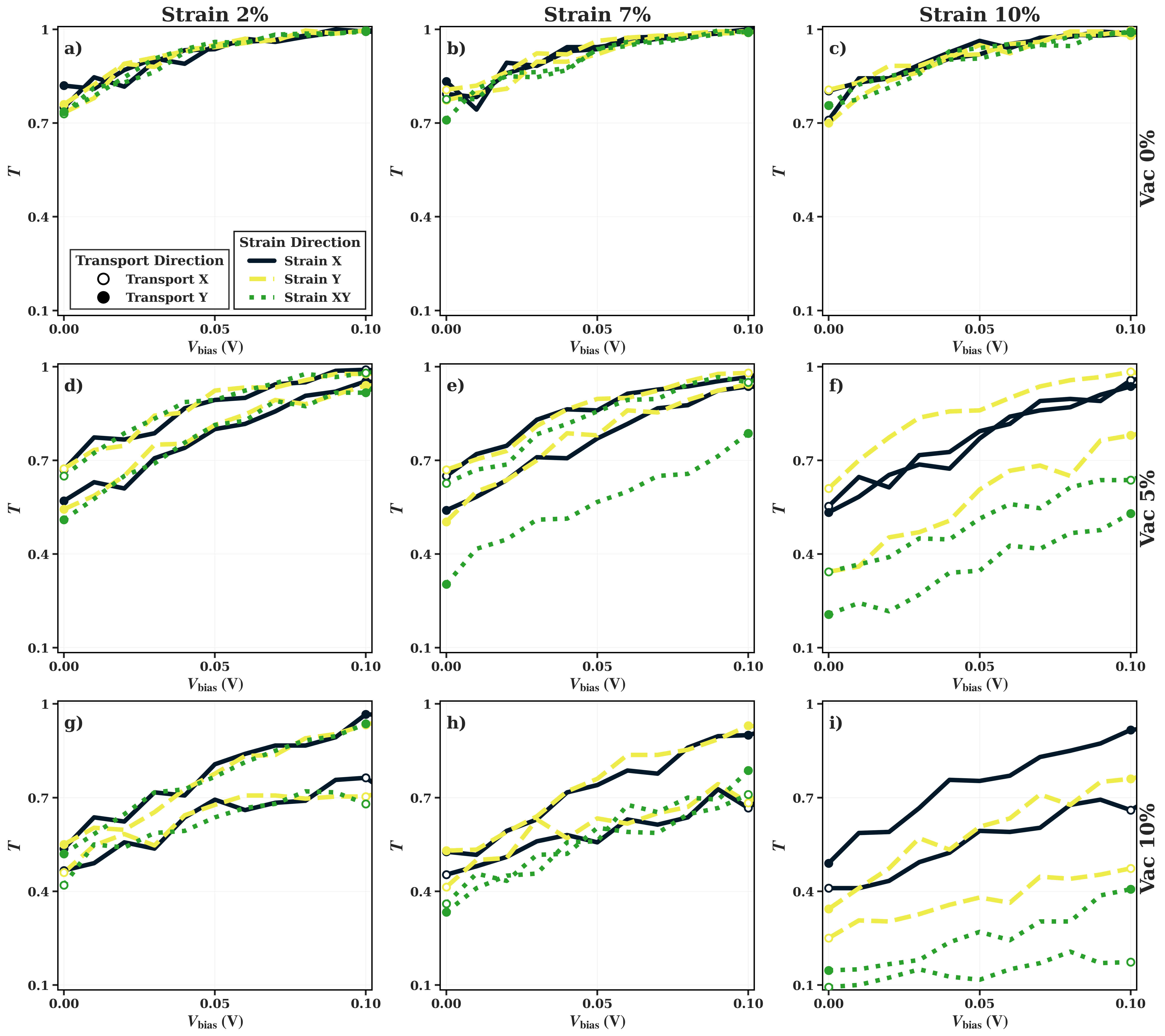}
\caption{Bias-dependent effective transmitance $T(V)$ of strained graphene with randomly distributed vacancies. Results are shown for vacancy concentrations of 0\%, 5\%, and 10\% (rows), strain magnitudes of 2\%, 7\%, and 10\% (columns), and three strain configurations: uniaxial strain along $X$, uniaxial strain along $Y$, and biaxial strain ($XY$). Transport is resolved along the $X$ and $Y$ directions, as indicated by open and filled symbols, respectively. Strain modifies transport both by altering inter-site distances in the hopping network and, in vacancy-containing systems, by enhancing localization through strain-induced changes in the effective transport gap. The strongest suppression of transmitance occurs when strain and vacancy concentration act cooperatively, particularly under biaxial strain and high disorder.}
\label{fig:fig5sup}
\end{figure}

\subsection*{Transport results for phagraphene}

To further assess the generality of the present transport framework, we applied the same methodology to phagraphene \cite{shi2021high, wang2015phagraphene}, a two-dimensional carbon allotrope with a lattice topology distinct from that of graphene. The goal of this supplementary analysis is not to provide an exhaustive characterization of phagraphene itself, but rather to verify that the main physical trends discussed for graphene are consistently reproduced in a structurally different sp$^2$ network. In this sense, phagraphene serves as an additional validation case for the proposed hopping-based approach.

As shown in the main text for graphene (see Figs.~\ref{fig:fig2}--\ref{fig:fig5}), the transport response results from the interplay between lattice connectivity, thermally activated hopping, magnetic-field-induced localization, and strain-modified geometry/electronic structure. The phagraphene results presented below exhibit the same overall phenomenology: temperature enhances transport by facilitating hopping, magnetic field suppresses transport through localization effects, and strain can modulate current depending on the transport direction and structural anisotropy. This qualitative agreement supports the robustness of the methodology and indicates that the framework captures general transport trends of disordered and perturbed two-dimensional carbon networks, rather than features specific only to graphene.

\begin{figure}[b!]
\centering
\includegraphics[width=1.0\linewidth]{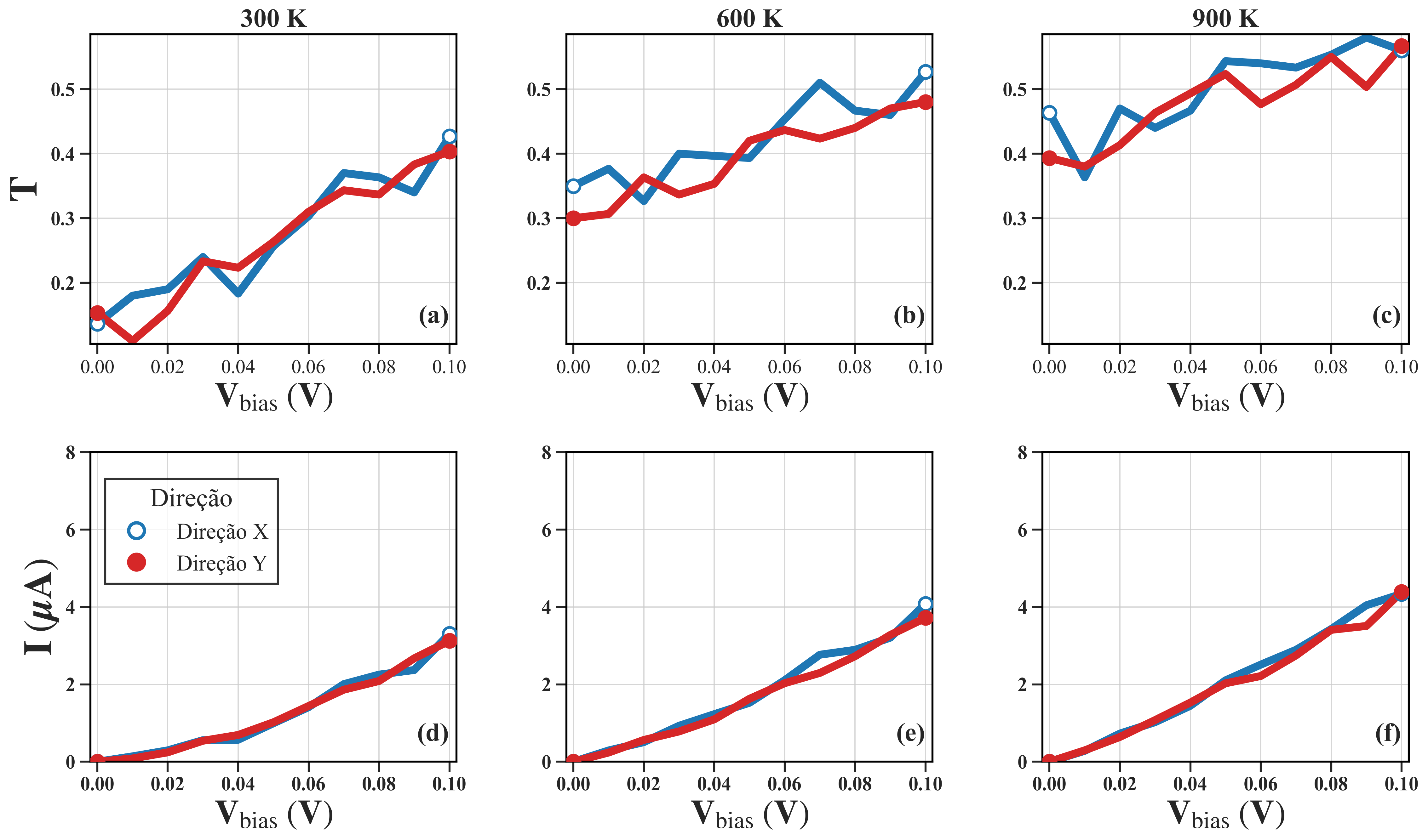}
\caption{Temperature-dependent transport response of phagraphene. The top panels show the effective transmittance $T(V)$ and the bottom panels show the corresponding current--voltage characteristics $I(V)$ for temperatures of 300 K, 600 K, and 900 K, resolved along the $X$ and $Y$ transport directions. Increasing temperature systematically enhances transmittance and current, consistent with thermally assisted hopping transport. Direction-dependent differences reflect the intrinsic anisotropy of the phagraphene lattice.}
\label{fig:phagraphene_temp}
\end{figure}

Figure~\ref{fig:phagraphene_temp} shows the temperature-dependent transport response of phagraphene. The top panels display the effective transmittance $T(V)$ and the bottom panels the corresponding current--voltage characteristics $I(V)$ for temperatures of 300, 600, and 900 K, resolved along the $X$ and $Y$ transport directions. As in graphene, increasing temperature systematically enhances both $T(V)$ and $I(V)$, consistent with thermally assisted hopping transport. At the same time, the differences between the $X$ and $Y$ directions are somewhat more evident, reflecting the intrinsic structural anisotropy of phagraphene. These results corroborate the interpretation adopted in the main text, namely that temperature primarily acts by increasing trajectory accessibility and hopping efficiency, while the detailed directional response is controlled by the underlying lattice topology.

\begin{figure}[b!]
\centering
\includegraphics[width=1.0\linewidth]{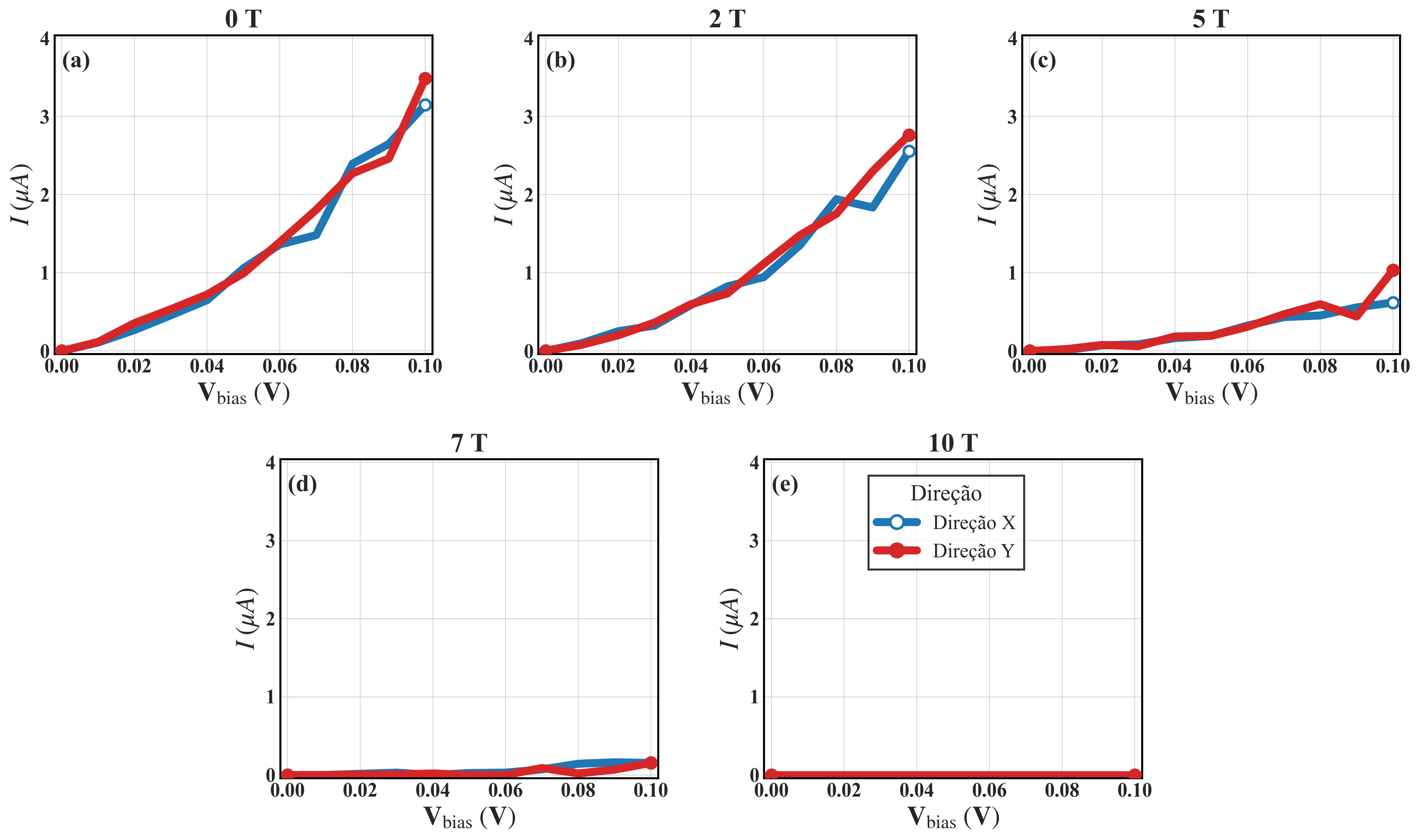}
\caption{Magnetic-field-dependent current--voltage characteristics of phagraphene. The current $I(V)$ is shown for magnetic fields of 0, 2, 5, 7, and 10 T, resolved along the $X$ and $Y$ transport directions. Increasing magnetic field systematically suppresses the current over the entire bias range, consistent with magnetic-field-induced localization and the consequent reduction of hopping efficiency. The stronger differences between the $X$ and $Y$ responses reflect the intrinsic structural anisotropy of the phagraphene lattice.}
\label{fig:phagraphene_bfieldI}
\end{figure}

Figure~\ref{fig:phagraphene_bfieldI} shows the magnetic-field-dependent current--voltage characteristics of phagraphene, for magnetic fields of 0, 2, 5, 7, and 10 T, with transport resolved along the $X$ and $Y$ directions. A systematic suppression of the current is observed as the magnetic field increases, with the response evolving from a finite and approximately linear $I(V)$ behavior at low field to an almost completely quenched transport regime at high field. This trend is fully consistent with the interpretation adopted in the main text for graphene, namely that the magnetic field reduces the effective localization length and progressively suppresses hopping conduction across the network. In phagraphene, however, the differences between the $X$ and $Y$ directions remain more evident, indicating that the intrinsic anisotropy of its lattice topology plays a stronger role in shaping the surviving percolation paths under magnetic confinement. These results therefore reinforce that the proposed framework captures transferable magnetotransport trends beyond graphene, while remaining sensitive to topology-dependent anisotropic effects.

\begin{figure}[b!]
 \centering
\includegraphics[width=1.0\linewidth]{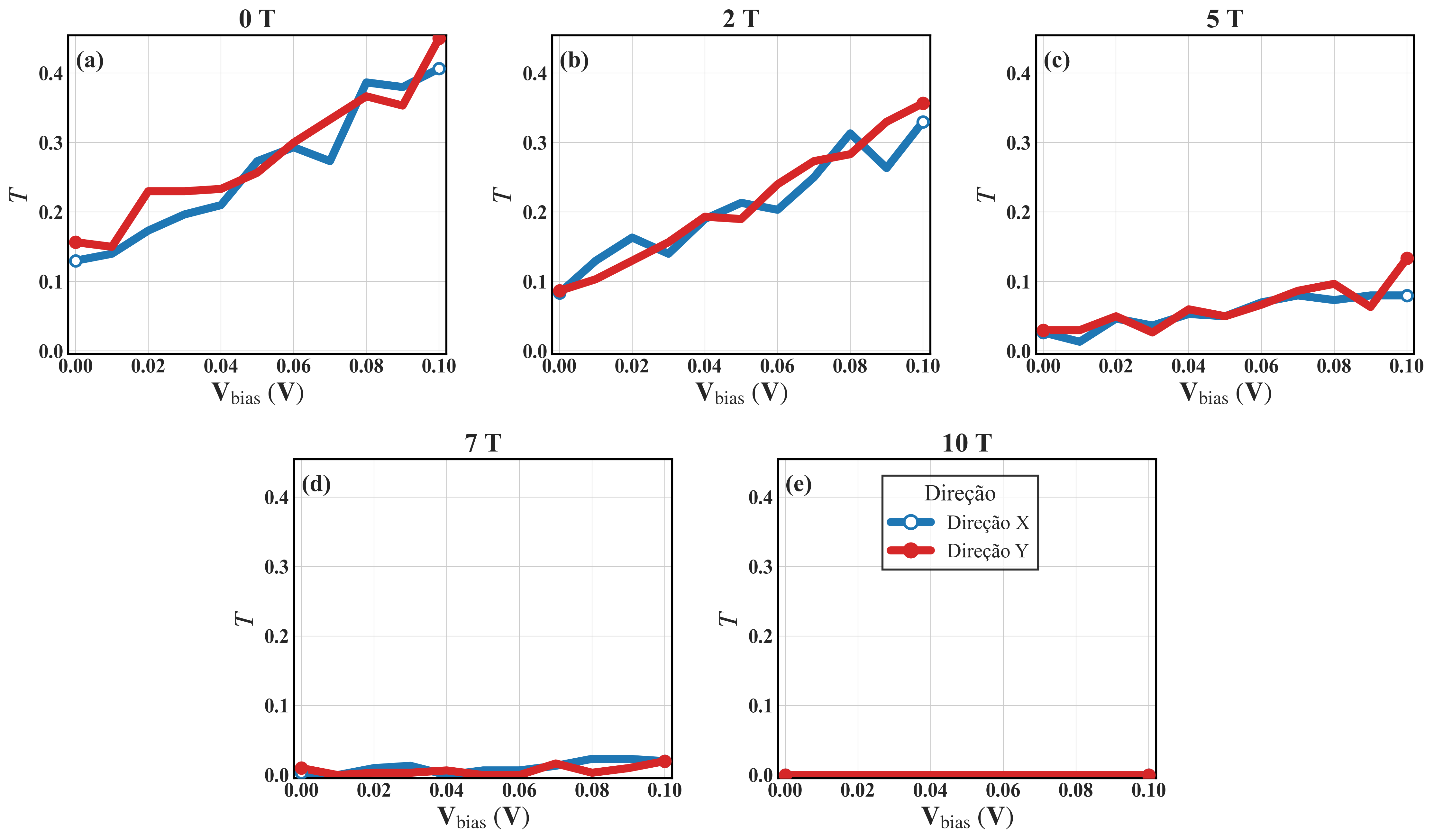}
\caption{Bias-dependent effective transmittance $T(V)$ of phagraphene under magnetic fields of 0, 2, 5, 7, and 10 T, resolved along the $X$ and $Y$ transport directions. Increasing magnetic field systematically suppresses the transmittance over the entire bias range, consistent with magnetic-field-induced localization and the resulting reduction in successful hopping trajectories. The directional differences between the $X$ and $Y$ responses reflect the intrinsic anisotropy of the phagraphene lattice.}
\label{fig:phagraphene_bfieldT}
\end{figure}

Figure~\ref{fig:phagraphene_bfieldT} shows the bias-dependent effective transmittance $T(V)$ of phagraphene under magnetic fields of 0, 2, 5, 7, and 10 T, with transport resolved along the $X$ and $Y$ directions. A progressive suppression of $T(V)$ is observed as the magnetic field increases, indicating that the fraction of successful carrier trajectories crossing the system is strongly reduced under magnetic confinement. This behavior is fully consistent with the interpretation adopted for graphene in the main text, where the magnetic field acts by reducing the effective localization length and therefore penalizing long or energetically marginal hopping paths. In phagraphene, the $X$ and $Y$ responses remain more clearly separated, revealing that the intrinsic anisotropy of its lattice topology has a direct impact on the stability of the percolation network under increasing magnetic field. The results therefore support the transferability of the proposed framework, showing that the same magnetically induced localization trends found in graphene are also reproduced in a structurally distinct two-dimensional carbon allotrope.

\begin{figure}[b!]
\centering
\includegraphics[width=1.0\linewidth]{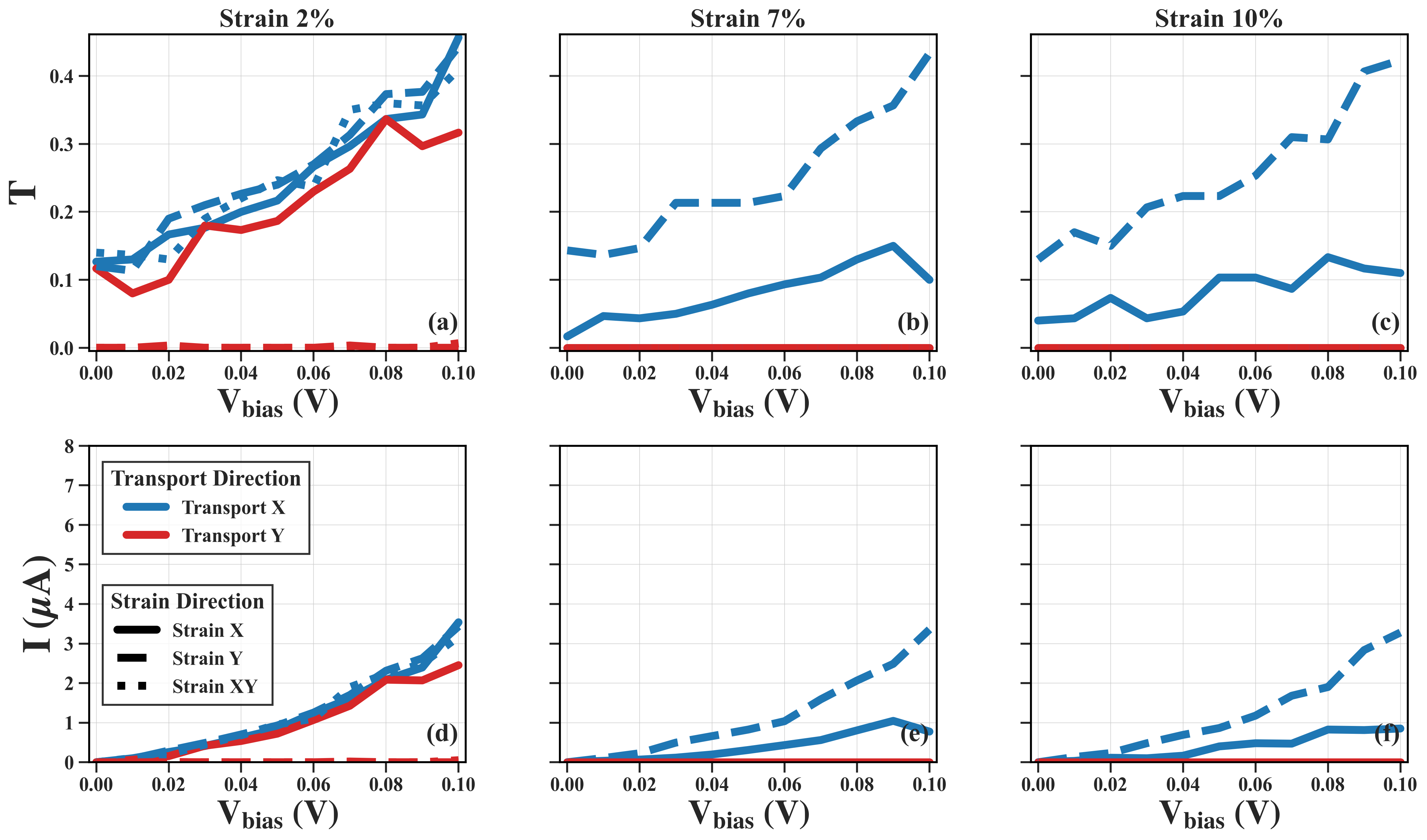}
\caption{Strain-dependent transport response of phagraphene. The top panels show the effective transmittance $T(V)$ and the bottom panels show the corresponding current--voltage characteristics $I(V)$ for strain magnitudes of 2\%, 7\%, and 10\%. Results are shown for uniaxial strain along $X$, uniaxial strain along $Y$, and biaxial strain ($XY$), with transport resolved along the $X$ and $Y$ directions. The pronounced dependence on strain orientation and transport direction reflects the intrinsic anisotropy of the phagraphene lattice and highlights the ability of the proposed framework to capture topology-dependent transport responses under mechanical deformation.}
\label{fig:phagraphene_strain}
\end{figure}

Figure~\ref{fig:phagraphene_strain} shows the strain-dependent transport response of phagraphene for strain magnitudes of 2\%, 7\%, and 10\%, considering uniaxial deformation along $X$, uniaxial deformation along $Y$, and biaxial strain ($XY$). The top panels display the effective transmittance $T(V)$ and the bottom panels the corresponding current--voltage characteristics $I(V)$, with transport resolved along the $X$ and $Y$ directions. In contrast to graphene, phagraphene exhibits a much stronger directional dependence under strain, with the transport response being highly sensitive both to the strain orientation and to the transport direction. In particular, strain applied along $Y$ strongly suppresses transport, whereas biaxial strain preserves a comparatively larger response along the $X$ direction, highlighting the role of the intrinsic lattice anisotropy in selecting the dominant percolation paths. These results are especially important from a methodological perspective: they show that the proposed framework is not limited to reproducing generic trends such as thermally assisted hopping or magnetic-field-induced suppression, but is also able to resolve topology-dependent anisotropic transport under mechanical deformation. The phagraphene case therefore provides a strong supplementary validation of the model, demonstrating that the method remains predictive and physically consistent when transferred to a structurally distinct two-dimensional carbon allotrope.

The phagraphene results presented in the Supplementary Material further support the main conclusion of this work: the proposed kinetic Monte Carlo hopping framework provides a physically consistent and transferable description of charge transport beyond the ballistic regime. In addition to reproducing the same qualitative trends observed for graphene under temperature, magnetic field, and strain, the method remains sensitive to the intrinsic anisotropy and topology of a structurally distinct carbon lattice. These results indicate that the framework is not restricted to a specific geometry, but can be extended to different two-dimensional carbon networks while preserving a clear connection between lattice structure, localization, and transport response.

\begin{acknowledgement}
 RMT acknowledges CNPq for financial support (grants 307371/2025-5 and 444069/2024-0). 
L.A.R.J. acknowledges financial support from FAPDF (grants 00193.00001808/2022-71 and 00193-00001857/2023-95), FAPDF-PRONEM (grant 00193.00001247/2021-20), PDPG-FAPDF-CAPES Centro-Oeste (grant 00193-00000867/2024-94), and CNPq (grants 350176/2022-1 and 167745/2023-9).
D.S.G. acknowledges support from the Center for Computing in Engineering and Sciences at Unicamp, through FAPESP/CEPID grant \#2013/08293-7.
\end{acknowledgement}


\bibliography{ref}

\end{document}